\documentclass{aa}  

\usepackage{graphicx}
\usepackage{txfonts}
\usepackage[T1]{fontenc}
\usepackage{subcaption} 
\usepackage{color}
\usepackage{amssymb}
\usepackage{epstopdf}
\usepackage{natbib}
\bibpunct{(}{)}{;}{a}{}{,} 
\usepackage{url}

\newcommand{\vsini}{$V \sin i$}   
 
\newcommand{\vmic}{$V_{\rm mic}$}
\newcommand{\vmac}{$V_{\rm mac}$}

\newcommand{\teff}{$T_{\rm eff}$}
\newcommand{\logg}{log\,{\it g$_\star$}}

\newcommand{\feh}{[Fe/H]}

\newcommand{\kms}{km\,s$^{-1}$}
\newcommand{\ms}{m~s$^{-1}$}

\newcommand{\gc}{g~cm$^{-3}$}
\newcommand{\lgr}{$\log\,(R^\prime_{HK})$}  

\newcommand{\Lsun}{L$_{\odot}$}                          
\newcommand{\Msun}{M$_{\odot}$}
\newcommand{\Rsun}{R$_{\odot}$}

\newcommand{\mearth}{M$_{\oplus}$}

\newcommand{\rearth}{R$_{\oplus}$}

\newcommand{\mstar}{$M_\star$}
\newcommand{\rstar}{$R_\star$}
\newcommand{\rhostar}{$\rho_{\mathrm{*}}$}

\newcommand{\rplanet}{$R_{\mathrm{p}}$}
\newcommand{\mplanet}{$M_{\mathrm{p}}$}

\newcommand{\targetaa}{EPIC\,220481411}
\newcommand{\target}{\mbox{K2-216}}

\newcommand{\bjdtdb}{\ensuremath{\rm {BJD_{TDB}}}}
 
\newcommand{\smassmean}[1][]{$0.70 \pm 0.03$ #1}   
\newcommand{\smasssp}[1][]{$0.70 \pm 0.03$ #1}   
\newcommand{\smassp}[1][]{$0.71 \pm 0.02$ #1}   
\newcommand{\smasssouth}[1][]{$0.74_{-0.09}^{+0.07}$ #1}   
\newcommand{\smassb}[1][]{$0.70 \pm 0.03$ #1}   
\newcommand{\smassspectral}[1][]{$0.71$ #1}   
\newcommand{\sradius}[1][]{$0.71\pm 0.07$ #1}  
\newcommand{\sradiusp}[1][]{$0.66\pm 0.02$ #1}  
\newcommand{\sradiussouth}[1][]{$0.77_{-0.18}^{+0.10}$ #1}   
\newcommand{\sradiusb}[1][]{$0.65 \pm 0.02$ #1}   
\newcommand{\sradiusspectral}[1][]{$0.70$ #1}   
\newcommand{\sradiusgaia}[1][]{$0.72\pm 0.03$ #1}  
\newcommand{\stempave}[1][]{$4\,503 \pm 69$ #1 } 
\newcommand{\sfehave}[1][]{$0.00 \pm 0.07$ #1} 
\newcommand{\sloggsp}[1][]{$4.57 \pm 0.09$ #1} 
\newcommand{\sloggp}[1][]{$4.63 \pm 0.02$ #1} 
\newcommand{\svsini}[1][]{$1.8 \pm 1.0$ #1} 
\newcommand {\lstargaia}[1][]{$0.19\pm 0.01$ #1 } 
\newcommand{\denspyaneti}[1][]{$2.3_{-1.4}^{+0.8}$ #1}  
\newcommand{\stype}{K5V} 
\newcommand{\age}{$5.0\pm 4.1$} 
\newcommand{\agebasta}{$8.2 _{-5.3}^{+4.8}$} 
\newcommand{\prot}{$30 \pm 5$} 
\newcommand{\distancegaia}{$115.8\pm1.5$} 
\newcommand{\parallaxgaia}{$8.6325\pm0.0525$} 
\newcommand{\velgaia}[1][$\mathrm{km\,s^{-1}}$]{ $-26.17 \pm 0.47$} 
\newcommand{\velHARPSN}[1][$\mathrm{km\,s^{-1}}$]{ $-26.0 \pm 0.1$} 
\newcommand{\velHARPS}[1][$\mathrm{km\,s^{-1}}$]{$-26.0 \pm 0.1$} 
\newcommand{\BCv}{$-0.64 \pm 0.02$} 

\newcommand{\Tzerob}[1][]{$7394.0417  \pm 0.0009$#1}  
\newcommand{\Pb}[1][]{$2.17480  \pm 0.00005$ #1}  
\newcommand{\bb}[1][]{$0.45  \pm 0.31$ #1}   
\newcommand{\arb}[1][]{$8.3 _{ - 2.3 } ^ { + 0.9}$ #1}   
\newcommand{\rrb}[1][]{$0.0221_{ - 0.0007  } ^ { +0.0022}$ #1} 
\newcommand{\ib}[1][]{$86.9 _{-4.0 }^{+2.2}$ #1}  
\newcommand{\ab}[1][]{$0.028 _{ - 0.007  } ^ {+ 0.003}$ #1} 
\newcommand{\rpb}[1]{$1.75  _{-0.10 }^{+0.17}$ #1}   
\newcommand{\Tequib}[1][]{$1103 _{ - 56 } ^ {+ 180}$  #1}    
\newcommand{\Fequib}[1][]{$248 _{ - 48  } ^ {+ 220}$  #1}    
\newcommand{\ttotb}[1][]{$1.84 _{ - 0.04} ^ { + 0.05}$  #1} 
\newcommand{\uone}[1][]{ $0.58  \pm 0.14$ #1}    
\newcommand{\utwo}[1][]{ $0.12   \pm 0.14$ #1}   

\newcommand{\kbold}[1][]{$3.8 _{-1.5}^{+1.3}$ #1}  
\newcommand{\kb}[1][]{$4.6 _{-1.4}^{+1.3}$ #1}  
\newcommand{\kbfco}[1][]{$5.0\pm1.0$}  
\newcommand{\mpb}[1]{$8.0  \pm 1.6$ #1}  
\newcommand{\mpbGP}[1]{$7.4  \pm 2.2$ #1}  
\newcommand{\mpbGPold}[1]{$6.1   _{-1.8}^{+1.6}$ #1}  
\newcommand{\hGP}[1]{$3.0^{+2.0}_{-1.2}$ #1}  
\newcommand{\hGPold}[1]{$2.4^{+1.6}_{-1.8}$ #1}  

\newcommand{\denpb}[1][]{$7.5 _{ - 2.6} ^ { + 3.0}$ #1} 
\newcommand{\denpbfco}[1][]{$8.2 _{ - 2.2} ^ { + 2.8}$ #1} 
\newcommand{\Lambdap}[1][]{$30$}  

\begin{document} 

\titlerunning{Super-Earth of 8~M$_\oplus$  in a 2.2-day orbit around the K5V star \target}
   \title{Super-Earth of 8~M$_\oplus$  in a 2.2-day orbit around the K5V star \target}

\author{C.~M.~Persson\inst{1} 
\and 
M.~Fridlund\inst{1,2}  
\and
O.~Barrag\'an\inst{3}  
\and
F.~Dai\inst{4, 5}  
\and
D.~Gandolfi \inst{3}   
\and
A.~P.~Hatzes\inst{6}
\and
T.~Hirano \inst{7}  
 \and 
S.~Grziwa \inst{8}  
\and 
J.~Korth \inst{8}
 \and
J.~Prieto-Arranz\inst{9,10}  
\and
L.~Fossati\inst{11}   
\and
V.~Van~Eylen\inst{2}
\and   
A.~B.~Justesen\inst{12}   
\and
J.~Livingston\inst{13}
\and
D.~Kubyshkina\inst{11}   
 \and 
H.~J.~Deeg\inst{9,10}
\and
E.~W.~Guenther\inst{6}
\and
G.~Nowak\inst{9,10}   
\and
J.~Cabrera \inst{14}
\and
Ph.~Eigm\"uller\inst{14}
\and
Sz.~Csizmadia\inst{14}
\and
A.~M.~S.~Smith\inst{14}  
 \and
A.~Erikson\inst{14}
%
\and
S.~Albrecht\inst{12}  
\and 
R.~Alonso~Sobrino\inst{9,10}
\and
W.~D.~Cochran\inst{15}  
\and
M.~Endl\inst{15}  
\and
M.~Esposito\inst{6}
\and
A.~Fukui\inst{16}
\and
P.~Heeren\inst{17}  
\and
D.~Hidalgo\inst{9,10}
\and
M.~Hjorth\inst{12}
\and
M.~Kuzuhara\inst{16,18}  
\and
N.~Narita\inst{13,16,18}  
\and
D.~Nespral\inst{9,10}
\and
E.~Palle\inst{9,10}   
 \and
M.~P\"atzold\inst{8}
 \and
H.~Rauer\inst{14,19}
\and
F.~Rodler\inst{20}  
\and
J.~N.~Winn\inst{5}  
 }  
    \offprints{carina.persson@chalmers.se}
   \institute{Chalmers University of Technology, Department of Space, Earth and Environment, Onsala Space Observatory,  SE-439 92 Onsala, Sweden.  
    \email{\url{carina.persson@chalmers.se}}
    \and
Leiden Observatory, University of Leiden, PO Box 9513, 2300 RA, Leiden, The Netherlands 
\and
Dipartimento di Fisica, Universit\'a di Torino, via Pietro Giuria 1, I-10125, Torino, Italy 
\and
Department of Physics and Kavli Institute for Astrophysics and Space Research, MIT, Cambridge, MA 02139, USA
\and
Department of Astrophysical Sciences, Princeton University, 024B, Peyton Hall, 4 Ivy Lane, Princeton, NJ 08544 
\and
Th\"uringer Landessternwarte Tautenburg,  D-07778 Tautenburg, Germany 
\and
Department of Earth and Planetary Sciences, Tokyo Institute of Technology, Meguro-ku, Tokyo, Japan 
\and
Rheinisches Institut f\"ur Umweltforschung an der Universit\"at zu K\"oln, Aachener Strasse 209, 50931 K\"oln, Germany 
\and
Instituto de Astrof\'isica de Canarias (IAC), 38205 La Laguna, Tenerife, Spain 
\and
Departamento de Astrof\'isica, Universidad de La Laguna, 38206 La Laguna, Tenerife, Spain 
\and
Space Research Institute, Austrian Academy of Sciences, Schmiedlstrasse 6, A-8042 Graz, Austria  
\and
Stellar Astrophysics Centre, Department of Physics and Astronomy, Aarhus University, Ny Munkegade 120, DK-8000 Aarhus C 
\and
Department of Astronomy, The University of Tokyo, 7-3-1 Hongo, Bunkyo-ku, Tokyo 113-0033, Japan 
\and 
Institute of Planetary Research, German Aerospace Center (DLR), Rutherfordstrasse
2, D-12489 Berlin, Germany 
\and
Department of Astronomy and McDonald Observatory, University of Texas at Austin, 2515 Speedway, Stop C1400, Austin, TX 78712, USA 
\and
National Astronomical Observatory of Japan, NINS, 2-21-1 Osawa, Mitaka, Tokyo 181-8588, Japan 
\and
Landessternwarte K\"onigstuhl, Zentrum f\"ur Astronomie der Universit\"at Heidelberg, K\"onigstuhl 12, 69117 Heidelberg, Germany  
\and
Astrobiology Center, NINS, 2-21-1 Osawa, Mitaka, Tokyo 181-8588, Japan  
\and
Center for Astronomy and Astrophysics, TU Berlin, Hardenbergstr. 36, 10623 Berlin, Germany  
\and
European Southern Observatory, Alonso de C\'ordova 3107, Vitacura, Casilla, 19001, Santiago de Chile, Chile  
}   

   \date{Received 21 Feb 2018; accepted 28 June 2018}

 
  \abstract
   {Although thousands of exoplanets have been discovered to date, far fewer have been fully characterised, in particular super-Earths. 
   The KESPRINT consortium identified   \target~as a  planetary candidate host star in the \emph{K2} space mission Campaign~8 field
   with a transiting super-Earth.  The planet has recently been validated as well.}
   {Our aim was to confirm the detection  and 
   derive the main physical characteristics of \target b, including the mass. } 
   {We performed a series of follow-up observations: high resolution imaging with the  
   FastCam camera at the TCS and the Infrared Camera and Spectrograph at Subaru,  
and high resolution spectroscopy with HARPS (La Silla), HARPS-N (TNG) 
and FIES (NOT). 
The stellar spectra were analyzed with the  
    {\tt{SpecMatch-Emp}} and {\tt{SME}} codes to derive the fundamental stellar  properties.
   We  analyzed  the \emph{K2} light curve with the {\tt{pyaneti}} software. 
    The   radial velocity   measurements were modelled with both a Gaussian process (GP) regression 
   and the so-called floating chunk offset (FCO)\ technique to simultaneously 
    model the planetary signal and correlated noise associated with stellar activity.}
   {Imaging confirms that \target~is a single star. Our  
   analysis discloses that the star is a 
   moderately active \stype~star of mass 
   \mbox{\smassmean~\Msun}~and radius \mbox{\sradiusgaia~\Rsun}. Planet~b is   found to have a radius of
   \mbox{\rpb~\rearth} and a 2.17-day orbit in agreement with previous results. 
We find consistent results for the planet mass from both models:    
   \mbox{\mplanet~$\approx$~\mpbGP~\mearth}  from the GP regression and 
    \mbox{\mplanet~$\approx$~\mpb~\mearth} from the FCO technique, 
    which implies  that this planet is a super-Earth. 
   The  incident stellar flux is \Fequib~F$_{\oplus}$. 
  }
   {The planet parameters  put planet b in the middle of, or just below,  the gap of the radius distribution
   of small planets. 
     The density  is consistent with 
     a rocky   composition of primarily iron and magnesium silicate. 
   In agreement with theoretical predictions, we find that the planet 
is a remnant core, stripped of its atmosphere,  and is one of the 
largest planets found that has lost its atmosphere. 
}


   \maketitle
%

\section{Introduction}
The NASA \emph{K2} mission \citep{Howell2014} is continuing the success of the \emph{Kepler} space mission 
by targeting stars in the ecliptic plane through high precision time-series photometry. 
Thousands of \emph{Kepler/K2}  exoplanet candidates have been discovered to date and hundreds have been confirmed and characterised. 
One of the surprises was the vast diversity of planets, in particular planets with radii between Earth and Neptune (3.9~\rearth), 
with no counterparts in our solar system.  
Short-period super-Earth planets, \rplanet~$\approx 1-1.75$~\rearth~\citep{2014ApJ...792....1L, 2017AJ....154..109F} have been
found to be very common based on planet occurrence rates and planet candidates discovered by \emph{Kepler} 
\citep{2015ApJ...809....8B}, 
although the number of 
well-characterised super-Earths are still low.  
Only a few dozen have both measured radius and mass\footnote{\url{https://exoplanetarchive.ipac.caltech.edu/}.}  
as of June 2018, and hence the composition and internal structure for the remaining super-Earths  are unknown. 

A bimodal radius distribution of small exoplanets at short orbital period was 
discovered by  \citet{2017AJ....154..109F}  using spectroscopic stellar parameters, and by 
\citet{2017arXiv171005398V} using asteroseismic stellar parameters. 
These findings show that very few planets at $P < 100$ days have sizes between 1.5 and 2~\rearth. 
The gap is predicted  by photo-evaporation models 
\citep{2013ApJ...776....2L, 2013ApJ...775..105O, 2014ApJ...795...65J, 2014ApJ...792....1L, 2016ApJ...831..180C, 2017ApJ...847...29O, 2018ApJ...853..163J}, 
in which close-in planets \mbox{($a < 0.1$~AU)} can lose their
entire atmosphere within a few hundred~Myr owing to intense stellar radiation. 
The mini-Neptunes   and super-Earths 
thus appear to be two distinct classes with radii of \mbox{$\sim2.5$~\rearth}, 
and \mbox{$\sim1.5$~\rearth}, respectively.  
These predictions, however, need to be tested against  well-characterised planets.

The work described in this paper is part of a larger programme performed by the  international 
	KESPRINT consortium\footnote{During 2016 the KESPRINT team was formed from  a merger of two teams: the "\emph{K2} Exoplanet Science Team" (KEST), 
and the "Equipo de Seguimiento de Planetas Rocosos Intepretando sus Transitos'' (ESPRINT) team; 
\tt{http://www.iac.es/proyecto/kesprint}.},  which combine \emph{K2} photometry with ground-based follow-up observations in order to confirm and characterise 
exoplanetary candidates \citep[e.g.][]{2017A&A...608A..93G, 2017AJ....153..131N,2017AJ....154..266N, 2018AJ....155..115L, 2018AJHirano, 2017AJ....153..130E, 2018MNRAS.474.5523S}. When processing  the  \emph{K2} Campaign~8 light curves, we found a super-Earth candidate
around \target~for which we proceeded with follow-up observations and characterisation described in this paper. 
During our work,  planet b was recently validated by \citet{2018AJ....155..136M}. 
In this paper, we confirm the planet and derive the previously unknown mass from radial velocity (RV) measurements.

The \emph{K2}  photometry  and transit detection are presented in 
Sect.~\ref{Section: K2 photometry and transit detection}. Ground-based follow-up observations (high resolution
imaging  and spectroscopy) are
presented in Sect.~\ref{Ground based follow-up}. We analyze the star in Sect.~\ref{Section: Stellar analysis} to obtain the 
necessary stellar mass and  radius for the transit analysis performed in Sect.~\ref{Section: transit analysis},  
and   the RV analysis carried out in Sect.~\ref{RV analysis}. 
We end the paper with a discussion and summary in
Sect.~\ref{Section: Discussion} and \ref{Section: Summary}, respectively.


%
\begin{table}[!t]
\caption{Main identifiers, coordinates, optical and infrared magnitudes, parallax and systemic velocity of  \target.}
\begin{center}
\begin{tabular}{lll} 
\hline\hline
     \noalign{\smallskip}
Parameter    & Value\tablefootmark{a}   \\
\noalign{\smallskip}
\hline
\noalign{\smallskip}
\multicolumn{2}{l}{\emph{Main Identifiers}} \\
\noalign{\smallskip}
EPIC & 220481411   \\
UCAC & 482-001110 & \\
2MASS & 00455526+0620490  \\
\noalign{\smallskip}
\hline
\noalign{\smallskip}
\multicolumn{2}{l}{\emph{Equatorial coordinates}} \\
\noalign{\smallskip}
$\alpha$(J2000.0) & $00^h\,45^m\,55\fs26$     \\
$\delta$(J2000.0) & 06$^{\circ}\,20\arcmin\,49\farcs10$     \\
\noalign{\smallskip}
\hline
\noalign{\smallskip}
\multicolumn{2}{l}{\emph{Magnitudes}} \\
$B$ (Johnson) & $13.563\pm0.020$      \\
$V$ (Johnson)  & $12.476\pm0.050$      \\
$Kepler$ & 12.10      \\
$g$ (Sloan) &  $13.043\pm0.030$     \\
$r$ (Sloan) &  $12.015\pm0.050$     \\
$i$ (Sloan) &  $11.696\pm0.010$     \\
$J$ (2MASS) &  $10.394\pm0.023$     \\
$H$ (2MASS)  &  $9.856\pm0.032$     \\
$K$ (2MASS)  &  $9.721\pm0.018$     \\
\noalign{\smallskip}
\hline
\noalign{\smallskip}  
Parallax   (mas) &\parallaxgaia\,$\tablefootmark{b}$  \\  
Systemic velocity (\kms) & \velgaia\,$\tablefootmark{b}$  \\    
    \noalign{\smallskip} \noalign{\smallskip}
\hline 
\end{tabular}
\tablefoot{
\tablefoottext{a}{All values  (except for Gaia DR2) are taken from the Ecliptic Plane Input Catalogue \citep[EPIC; ][]{2016ApJS..224....2H} available at \url{http://archive. stsci.edu/k2/epic/search.php}.} 
\tablefoottext{b}{Gaia~DR2; \url{http://gea.esac.esa.int/archive/}.}
}
\end{center}
\label{Table: Star basic parameters}
\end{table}

  \begin{figure*}[!ht]
 \centering
  \resizebox{\hsize}{!}
            {\includegraphics{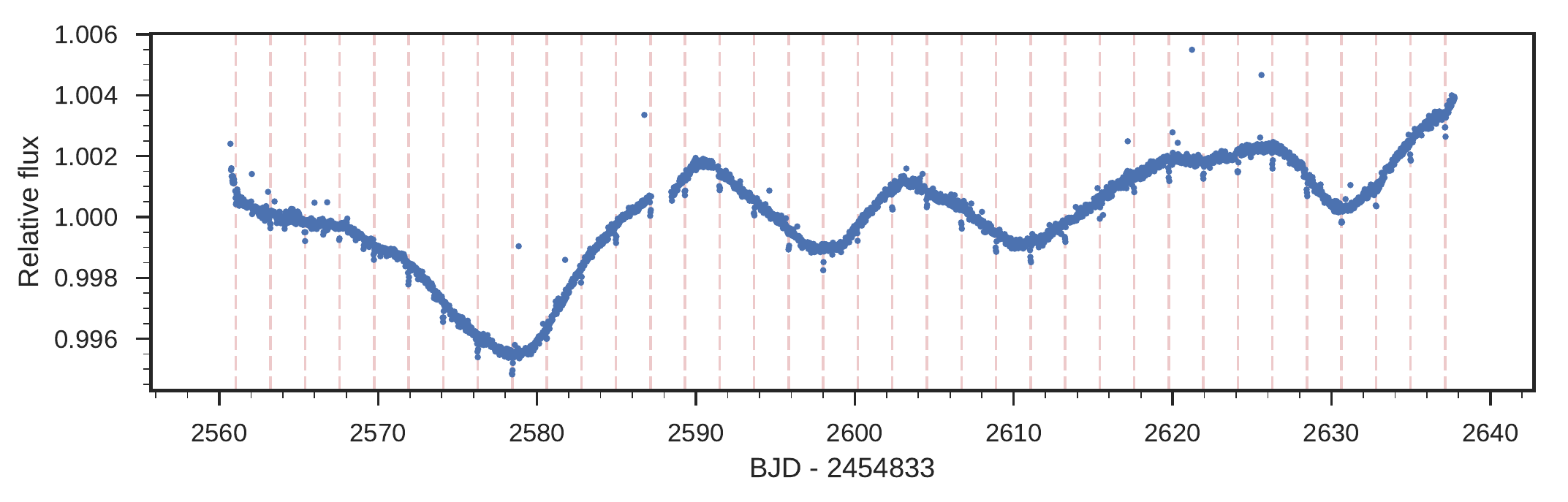}}
   \caption{Full pre-processed  light curve of \target~by Vanderburg.  The stellar activity is seen as the long period modulation. 
   The narrow and shallow 36 planet transits used in the analysis are  indicated with dotted vertical lines. }
      \label{Figure: full lightcurve}
 \end{figure*}

 \section{\emph{K2} photometry and transit detection} \label{Section: K2 photometry and transit detection}
Observations of the \emph{K2} Field~8 took place from Jan 4 to March 23, 2016.  
The telescope was pointed at the coordinates $\alpha = 01^h05^m21^s$ and $\delta = +05^\circ15\arcmin44\arcsec$  ($J2000$).  
A total of 24\,187 long-cadence (29.4 min integration time) and 55 short-cadence (1~min integration time) targets 
were observed. 

We downloaded the \emph{K2} Campaign~8 data  
from   the Mikulski Archive for 
Space Telescopes\footnote{\url{https://archive.stsci.edu/k2/epic/search.php}} (MAST). 
For the detection of transiting candidates,
we searched   the data 
using three different 
methods,  optimised for space-based photometry: 
(\emph{i}) the {\tt{EXOTRANS}} \citep{Grziwa2012}   
routines, (\emph{ii}) the  
D\'etection Sp\'ecialis\'ee de Transits  ({\tt{DST}})  software  \citep{Cabrera2012}, and  (\emph{iii})   
a   method similar to that described by \citet{2014PASP..126..948V}.  
The codes have been used extensively 
on \emph{CoRoT}, \emph{Kepler} and other \emph{K2} campaigns.  
The strategy of using different software has   been shown 
to be successful, since   both the false alarm and non-detections are model dependent. 

{\tt{The EXOTRANS}}  and {\tt{DST}} methods were applied to the   pre-processed light curves by Vanderburg 
 using the method described in
\citet{VandJohn2014}.  
 {\tt{The EXOTRANS}}  method is built on a combination of the wavelet-based filter technique {\tt{VARLET}}  \citep{2016arXiv160708417G}
 and a modified version of the {\tt{BLS}} \citep[Box-fitting Least Squares; ][]{Kovacs2002} 
 algorithm to detect the most significant transit. 
 When a significant transit is detected, the {\tt{Advanced BLS}} removes a detected transit 
 using a second wavelet based filter routine known as  {\tt{PHALET}}. This routine combines phase-folding and 
 wavelet based approximation to recreate and remove periodic features in light curves. 
 After removing a detected transit, the light 
 curve is searched again for additional transits.  This process is repeated 15 times to 
 detect multi-planet systems. Since the detected features are completely removed,  
 transits near resonant orbits are easily found. The
 {\tt{DST}}  method aims at a specialised detection of transits by improving the consideration of the transit shape 
and the presence of transit timing variations. The same number of free parameters as {\tt{BLS}} are used,  
and the code  implements
better statistics with signal detection. 
In the third method,  described in more detail by \citet{2016ApJ...823..115D} and
 \citet{2018AJ....155..115L}, we extracted aperture photometry and image centroid 
position information   from the \emph{K2} pixel-level data to decorrelate the flux variation from the
 rolling motion of the telescope  to produce our own light curves.  The transit detection routine   utilises the standard {\tt{BLS}} 
 routine   and an optimal 
 frequency sampling \citep{2014A&A...561A.138O}.

A shallow transit signal was discovered by all three methods in the light curve of  \target~(\targetaa) with
a period of \mbox{$\sim2.2$~days} and  a depth of $\sim0.05\ \%$ consistent with a super-Earth orbiting a \stype~star. 
We searched for even-odd transit depth variation and secondary eclipse, which would point
to a binary scenario, but neither were detected within $1\sigma$. 
\target~ was   proposed by programme GO8042 and observed in the long-cadence mode. 
The basic parameters of the star are listed in Table~\ref{Table: Star basic parameters}.
The full pre-processed   light curve by Vanderburg\footnote{Publicaly available at  \url{https://www.cfa.harvard.edu/~avanderb/k2c8/ep220481411.html}}    is shown in Fig.~\ref{Figure: full lightcurve} in which  
36 clear  transits are denoted with dotted vertical lines.

\section{Ground-based follow-up} \label{Ground based follow-up}
Follow-up observations 
were performed to determine whether the signal is from a planet 
and  to obtain further information on the planet properties. 
High resolution imaging was used to check if the transit is a false positive from
a fainter unresolved binary  
included in the  \emph{K2} sky-projected pixel size of $\sim 4$\arcsec~in 
\mbox{Sect.~\ref{Subsubsection: Fastcam imaging} -- \ref{Subsubsection: speckle imaging}.}   
The presence of a  
binary can   lead to an erroneous  radius of the transiting object, which propagates into the  
density;  this is important
 for distinguishing between rocky planets
and those with an envelope (mini-Neptunes). The binary can be either
an unrelated background system or a companion to the primary star. 
The planetary nature of the transit  was then confirmed by our   
high resolution RV measurements described in Sect.~\ref{Section: RV measurements}, 
which also allows a measure of its mass (Sect.~\ref{RV analysis}). 
 This data was also used
to derive stellar fundamental parameters with spectral analysis codes (Sect.~\ref{Section: Stellar analysis}).

\subsection{FastCam imaging and data reduction} \label{Subsubsection: Fastcam imaging}

\begin{figure}[!ht]
 \centering
\includegraphics[width=0.41\textwidth]{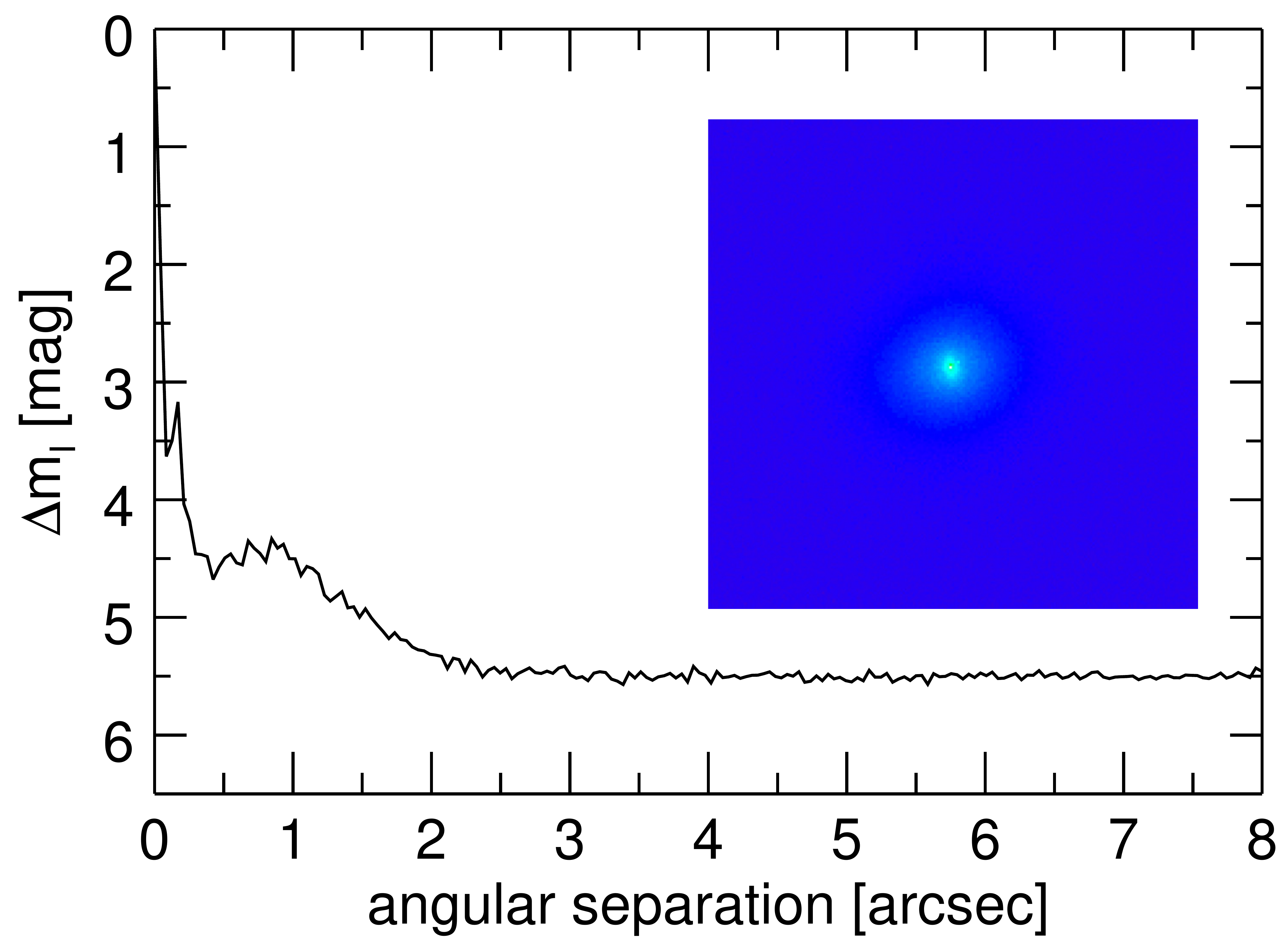}
   \caption{\emph{I} band  magnitude $5\sigma$ contrast curve as a function of angular 
   separation up to  8\arcsec~from   \target~obtained with the FastCam camera at TCS. 
 The inset shows the $8\arcsec \times 8\arcsec$    image.}
      \label{Figure: lucky Jorge imaging}
 \end{figure}

We performed Lucky Imaging (LI) of \target~with the FastCam camera 
\citep{Oscoz2008} at 1.55 m  Telescopio Carlos S\'anchez (TCS). The FastCam 
is a very low noise,  high-speed electron-multiplying charge-coupled device (CCD)  camera with $512 \times 
512$ pixels, a physical pixel size of 16 microns, and a field of view of 
$21\farcs2\times21\farcs2$. On the night of Sept 6 (UT), 2016, $10\,000$ 
individual frames of \target~were collected in the Johnson-Cousins infrared 
\mbox{\emph{I} band} (880~nm) with an exposure time of 50 ms for each frame. The 
typical Strehl ratio in our observation varies with the percentage of 
the best-quality frames chosen in the reduction process as follows: from 0.05 for  the
 90~\% to  0.10  for   1~\% best images. In order to construct a high 
resolution, long-exposure image, each individual frame was 
bias-subtracted, aligned and co-added, and then processed with the 
FastCam dedicated software developed at the Universidad Polit\'ecnica de 
Cartagena \citep{2010SPIE.7735E..0XL, 2013MNRAS.429..859J}. 
The inset in Fig.~\ref{Figure: lucky Jorge imaging} shows a high 
resolution image, which was constructed by co-addition of the 30~\% 
best images, with a 150~s total exposure time.  Figure~\ref{Figure: lucky Jorge imaging} also   
shows the $5\sigma$ contrast curve, which quantitatively describes the detection limits of nearby possible companions that are 
computed based on the scatter within 
the annulus as a function of angular separation from the target 
centroid \citep{2017A&A...597A..47C}.  As shown by the contrast curve, no bright companion was detectable  within 8\arcsec. 
Between 2\arcsec~and 8\arcsec~separation we can exclude companions   $\approx6\times10^{-3}$  times brighter than \target.

\subsection{Subaru/IRCS AO imaging and data reduction} \label{Subsubsection: AO imaging}
In order to further check for possible unresolved eclipsing binaries mimicking planetary transits, we imaged \target~with the
Infrared Camera and Spectrograph \citep[IRCS;][]{2000SPIE.4008.1056K}
with the adaptive optics (AO) system \citep{2010SPIE.7736E..0NH} on the 
Subaru 8.2 m telescope producing diffraction  limited images in the $2-5\ \mu$m range. 

  \begin{figure}[!ht]
   \centering
\includegraphics[width=0.41\textwidth]{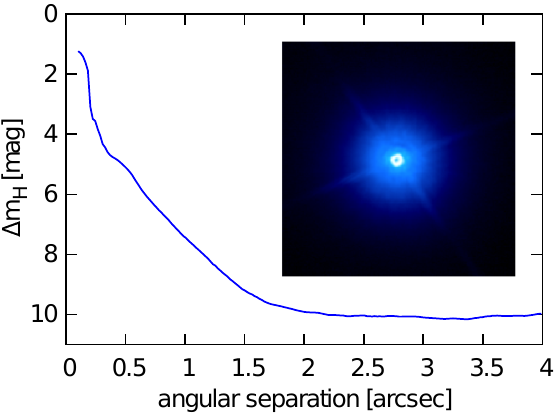}
   \caption{\emph{H} band (1630~nm)  $5\sigma$ magnitude contrast curve as a function of angular separation from   
   \target~obtained with  IRCS/Subaru.  
The inset shows the $4\arcsec \times 4\arcsec$   saturated   image.}
      \label{Figure: subaru Teru imaging}
 \end{figure}

  \begin{figure}[!ht]
\hspace{1.7em}
\includegraphics[width=0.43\textwidth, height=5.9cm]{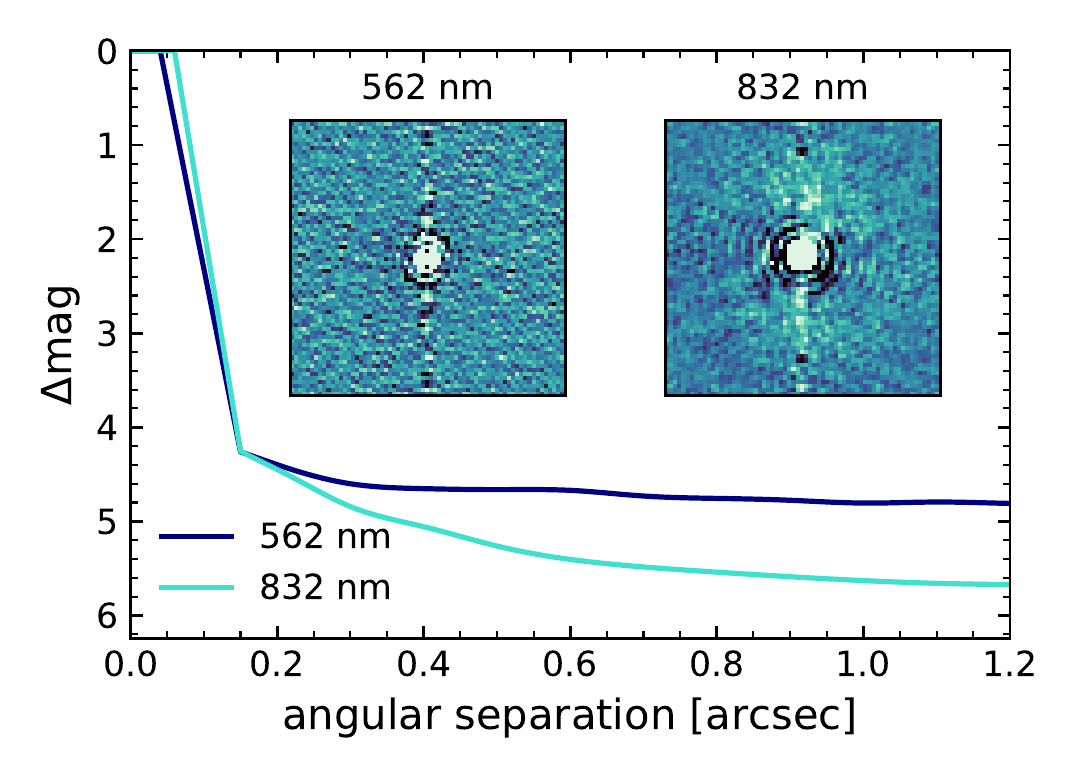}
   \caption{Reconstructed images in the $r$-  and  $z$-narrowbands  from NESSI/WIYN speckle interferometry and the resulting 
  $5\sigma$   contrast curves. The inset images are $1\farcs2\times1\farcs2$. Northeast is up and to the left.}
      \label{Figure: speckle imaging}
 \end{figure}

The high resolution mode was selected at a pixel scale of 0\farcs0206~per pixel,
and a field of view of 21\arcsec$\times\,21$\arcsec. 
Adopting \target~itself as a natural guide 
star, we performed AO imaging on Nov 6, 2016 in the \emph{H} band (1630~nm) with two different exposures.  
The first 
sequence consists of a short exposure \mbox{($0.4~\mathrm{s} \times 3$ co-additions)}  
with the five-point dithering to obtain unsaturated target images  
for the absolute flux calibration. We then 
repeated longer exposures \mbox{($5~\mathrm{s} \times 3$ co-additions)} with 
the same five-point dithering for saturated images to look for faint nearby 
companions. The total scientific exposure time amounted to 225~s. 
We reduced the IRCS AO data following \citet{2016ApJ...820...41H}; 
we applied the dark subtraction, flat fielding, distortion correction, and 
aligned the frames, which were subsequently median-combined to obtain
the final   images for unsaturated and saturated frames, respectively. 

We found that the  full width 
at half maximum (FWHM)  of \target~is $0\farcs096$, 
as measured from the combined unsaturated image. 
The inset of Fig.~\ref{Figure: subaru Teru imaging} shows the combined saturated image with 
a field of view of \mbox{$4\arcsec \times 4\arcsec$}. 
To estimate the contrast 
achieved by the IRCS imaging, we convolved the combined saturated image 
with the target's  FWHM  and computed the scatter within the small annulus 
centred at the centroid of the target. The $5\sigma$ 
contrast curve as a function of angular separation from the target is     
drawn in Fig.~\ref{Figure: subaru Teru imaging}.  
No bright nearby sources were found around \target. 
For instance, the contrast curve shows that at a separation of 0\farcs5~(1\farcs0), companions 
brighter than $\Delta m_H\sim5$~mag ($\sim7.5$~mag) 
would have been detected with $>5\sigma$. Thus we can
exclude companions brighter than $1\times10^{-3}$  of the target star at a 
separation of 1\arcsec.

\subsection{NESSI imaging} \label{Subsubsection: speckle imaging}
For comparison with our FastCam and IRCS imaging, we also show speckle imaging   of \target~performed with the   
NASA Exoplanet Star and Speckle Imager  \citep[NESSI;][Scott et~al. in prep.]{2016SPIE.9907E..2RS} 
at the WIYN 3.5 m telescope in Fig.~\ref{Figure: speckle imaging}. The images were retrieved from ExoFop\footnote{\url{https://exofop.ipac.caltech.edu/k2/.}} 
(with permission from the observers). The contrast curves based on the 
same data were also used in \citet{2018AJ....155..136M} to calculate the false-positive probabilities  (FAP).  
The observations were conducted at 562~nm \mbox{($r$-narrowband)} and 832~nm \mbox{($z$-narrowband)} simultaneously
on Nov  14, 2016. 
The data were collected 
and reduced following the procedures described by \citet{2011AJ....142...19H}. 
The resulting reconstructed 
images of the host star are $4\farcs6\times4\farcs6$ with a resolution close to the diffraction 
limit of the telescope ($0\farcs040$ at 562~nm and $0\farcs060$ at 832~nm). 
No secondary sources were detected in the reconstructed images. We  produced the
$5\sigma$ detection limits from the reconstructed images using a series of concentric 
annuli as shown up to 1\farcs2~in Fig.~\ref{Figure: speckle imaging}.

\begin{table*}[!t]
 \centering
 \caption{FIES, HARPS, and HARPS-N RV measurements of  \target.}
 {\begin{tabular}{lcccccccccc}
\hline
\hline
\noalign{\smallskip}
\bjdtdb\tablefootmark{a}&   \multicolumn{2}{c}{RV\tablefootmark{b}  (\kms) } & $t_\mathrm{exp}$ & S/N\tablefootmark{c} &  GP$\tablefootmark{d}$  & FCO\tablefootmark{e}  &
BIS\tablefootmark{f}  &  FWHM\tablefootmark{g} 
&  \multicolumn{2}{c}{\lgr\tablefootmark{h} }\\
(-2\,450\,000.0) & Value & Error & (s) & Value &  Y/N   & ``chunk'' & (\kms) & (\kms) & Value & Error \\
\noalign{\smallskip}
\hline
\noalign{\smallskip}
FIES \\
($\gamma= +1.1\pm4.5$~\ms)\,\tablefootmark{i} \\
\noalign{\smallskip}
7640.651587  &   0.0000     &         0.0049&3\,600      &\ldots&        Y & 1&\ldots& \ldots &\ldots& \ldots\\ 
7641.696953  &   0.0026     &         0.0055&3\,600       &\ldots&       Y&1&\ldots& \ldots &\ldots& \ldots\\ 
7642.604136  &  -0.0078     &         0.0064&3\,600       &\ldots&       Y&1&\ldots& \ldots &\ldots& \ldots\\  
7643.653029  &   0.0097     &         0.0060&3\,600       &\ldots&       Y&1&\ldots& \ldots &\ldots& \ldots\\  
\noalign{\smallskip}
FIES2 (new CCD) \\
($\gamma= -4.2\pm4.4$~\ms)\,\tablefootmark{i} \\
\noalign{\smallskip}
7682.486821  &  0.0000 &             0.0064&3\,600         &\ldots&      Y&2&\ldots& \ldots &\ldots& \ldots\\  
7683.631236  & -0.0089 &             0.0067&3\,600         &\ldots& Y&2&\ldots& \ldots &\ldots& \ldots\\ 
7684.494008  & -0.0034 &             0.0065&3\,600         &\ldots& Y&2&\ldots& \ldots &\ldots& \ldots\\ 
7717.444472  & -0.0040 &             0.0093&3\,600         &\ldots& N&\ldots&\ldots& \ldots &\ldots& \ldots\\ 
\noalign{\smallskip}
HARPS  \\
($\gamma = -25903.7\pm2.2$~\ms)\,\tablefootmark{i} \\
\noalign{\smallskip}
7682.680755  &     -25.8996  &     0.0020  &3\,600      &45.5   &  Y    &\ldots &       0.055   &         6.30    &       -4.675  &  0.017\\ 
7984.899412  &     -25.8984  &     0.0027  &3\,600      &34.3   &  Y    &3      &       0.041   &         6.27    &       -4.659  &        0.023\\ 
7985.874006  &     -25.9063  &     0.0033  &3\,600      &28.6   &  Y    &3      &       0.049   &       6.26         &       -4.621 &         0.024  \\ 
7986.881922  &     -25.9027  &     0.0024  &3\,600      &37.0   &  Y    &3/4    &       0.058   &       6.27  &      -4.665&  0.019  \\ 
7987.861213  &     -25.9036  &     0.0019  &3\,600      &46.0   &  Y    &4      &       0.050   &       6.27    &       -4.655&  0.014    \\ 
7990.892233\tablefootmark{j}  &     -25.9141  &     0.0055  &3\,600     &19.3   &  N      &\ldots &       0.049   &       6.31    &       -4.632&  0.044   \\ 
7991.870060  &     -25.9036  &     0.0030  &3\,600      &30.6   &  Y    &4      &       0.050   &         6.27    &       -4.624 &         0.024  \\ 
8003.765390  &     -25.9052  &     0.0035  &3\,600      &27.7   &  Y    &\ldots &       0.033   &       6.28    &       -4.752  &  0.025        \\ 
8082.577859  &     -25.9108  &     0.0025  &3\,600      &35.3   &  Y    &\ldots &       0.048   &         6.27    &       -4.679  &  0.018        \\ 
\noalign{\smallskip}
HARPS-N  \\
($\gamma = -25910.3\pm1.8$~\ms)\,\tablefootmark{i} \\
\noalign{\smallskip}
7692.420875  &   -25.9157  &   0.0024  &2\,700  &34.5   &  Y  &5&       0.035   &         6.22    &       -4.721  &        0.019  \\
7693.429280  &   -25.9071  &   0.0028  &2\,700  &32.9   &  Y    &5&     0.036   &       6.23         &       -4.676  &        0.020  \\ 
7694.406191  &   -25.9130  &   0.0026  &2\,700  &32.6   &  Y    &5&     0.037   &       6.22         &       -4.688  &        0.019  \\ 
7694.559784  &   -25.9216  &   0.0027  &2\,700  &33.1   &  Y&5& 0.039   &       6.23    &       -4.693  &        0.021   \\ 
7743.412714  &   -25.9153  &   0.0046  &3\,300  &22.8   &  N    &\ldots&        0.032   &       6.25         &       -4.632  &        0.035  \\ 
8013.524085  &   -25.9034  &   0.0019  &3\,600  &44.7   &  Y    &6&     0.047   &       6.29         &       -4.625  &        0.011  \\
8013.705318  &   -25.9013  &   0.0025  &3\,600  &37.7   &  Y    &6& 0.046               &       6.29    &       -4.618  &        0.014   \\ 
8014.548213  &   -25.9137  &   0.0034  &3\,600  &29.1   &  Y    &6&     0.053   &       6.29         &       -4.623  &        0.022  \\ 
8046.512566  &   -25.9070  &   0.0028  &3\,360  &33.4   &  N    &\ldots& 0.044           &       6.29    &       -4.644  &        0.018  \\
8054.600828\tablefootmark{j}   &   -25.9118  &  0.0059 &2\,865  &19.6   &  N      &       \ldots& 0.023           &      6.28 &      -4.620       &        0.044      \\
8080.406242  &   -25.9122  &   0.0021  &3\,300   &40.0  &  Y    &\ldots &0.037          &       6.26         &       -4.681      &    0.014  \\
8109.383188\tablefootmark{j}   &   -25.9169  &          0.0058 &3\,360  &19.5   & N       &\ldots & 0.028         &       6.25 &  -4.681  &        0.050  \\
8129.370822 &    -25.9028  &    0.0037 &3\,360  &27.4           &  N&\ldots     & 0.043           & 6.28 &  -4.634  &        0.026  \\
\noalign{\smallskip}
\hline
\end{tabular}
}
\tablefoot{
\tablefoottext{a}{Time stamps are given in barycentric Julian day in barycentric dynamical time.} 
\tablefoottext{b}{The FIES RV measurements are relative, while the HARPS and HARPS-N measurements
are absolute.} 
\tablefoottext{c}{S/N per pixel at 5500\,\AA.}
\tablefoottext{d}{Included RVs in the GP regression  model.}   
\tablefoottext{e}{The division of chunks in the  FCO technique  RV model. The
model excluded the isolated RVs in the empty entries.}   
\tablefoottext{f}{Bisector inverse slope      of the CCF.}   
\tablefoottext{g}{FWHM of the CCF.} 
\tablefoottext{h}{A dimensionless ratio of the emission in the \ion{Ca}{II} H and K line cores
to that in two nearby continuum bandpasses on either side of the  lines.}
\tablefoottext{i}{Relative (FIES) and absolute (HARPS and HARPS-N) systemic velocities derived from the Gaussian regression analysis.}
\tablefoottext{j}{Not used in the any of the RV models owing to \mbox{S/N $< 20$}.}
}
 \label{Table: RV measurements}
\end{table*}

\begin{table*}[!htbp] 
 \centering
 \caption{Spectroscopic parameters of \target~as derived from the co-added HARPS and HARPS-N spectra using
 {\tt {SME}} and  {\tt {SpechMatch-emp}}.}
\begin{tabular}{lccccc}
 \hline\hline
     \noalign{\smallskip}
 ~~&   ~~~~~~~~~~~~~~~~~~~~~~~~~~~~~~~ \teff~~~~~~~~~~~ &         \logg~~~~~~         &       ~~~~~~~~[Fe/H]  &       ~~~~~~~~ [Ca/H]    &     ~~~~\vsini~~         \\
~~ &   ~~~~~~~~~~~~~~~~~~~~~~~~~~~~~~~ (K)~~~~~~~~~~~      &       (cgs)~~~~~~         &      ~~~~~~~~(dex)    &       ~~~~~~~~   (dex)     &  ~~~~(\kms)~  \\
    \noalign{\smallskip}
     \hline
\noalign{\smallskip} 
\end{tabular}
\begin{tabular}{lllrcc}
HARPS \\
 {\tt {SME}}  
 & $4\,520\pm136$       &        $4.33\pm0.20$  &       $-0.05 \pm 0.11$                &       $-0.09 \pm 0.10$       & \svsini\\
 {\tt {SpecMatch-Emp}}   
 &$4\,426\pm70$ &       $4.58\pm0.09\,\, \tablefootmark{a}$     &       $0.05 \pm 0.12$& \ldots       & \ldots  \\
    \noalign{\smallskip}
     \hline
\noalign{\smallskip} 
HARPS-N \\
 {\tt {SME}}  &  $4\,500 \pm 140$       &        $4.37\pm0.20$  &       $0.00 \pm 0.12$    &  $-0.08 \pm 0.10$                & \svsini\\
 {\tt {SpecMatch-Emp}} 
 &$4\,490\pm 70$        &        \sloggsp$\tablefootmark{a}$    &       $0.06 \pm 0.12$        & \ldots       & \ldots \\
    \noalign{\smallskip} \noalign{\smallskip}
\hline 
\end{tabular}
\tablefoot{
\tablefoottext{a}{Coupling the {\tt{SpecMatch-Emp}}  modelling  with  the calibration equations from  \citet{Torres2010}.}
}
 \label{Table: Spectroscopic parameters}
\end{table*}

\subsection{High resolution spectroscopy} \label{Section: RV measurements}

We performed high resolution spectroscopy  
to obtain  RV measurements with three different instruments:    FIES,   HARPS, and  HARPS-N. 
All RVs are listen in Table~\ref{Table: RV measurements}. 

\emph{FIES}: 
 We started the RV follow-up of \target~with  the FIbre-fed \'Echelle 
Spectrograph \citep[FIES;][]{Frandsen1999, Telting2014} mounted at the 
2.56 m Nordic Optical Telescope (NOT) of Roque de los Muchachos 
Observatory (La Palma, Spain). Eight high resolution spectra 
($R\,\approx\,67\,000$) were gathered between Sept and Nov 
2016, as part of our \emph{K2} follow-up programmes 53-016, 54-027, and 
54-211. To account for the RV shift caused by the replacement of the 
CCD, which occurred on 30~Sept~2016, we treated 
the spectra taken in Sept~2016 and those acquired in 
Oct--Nov~2016 as two independent data sets. We set the exposure 
time to 3600~s and followed the same observing strategy described in 
\citet{Gandolfi2013} and \citet{Gandolfi15}, i.e. we traced the RV drift of the instrument by 
bracketing the science exposures with long-exposed ThAr spectra. 
The data reduction was performed using standard IRAF and IDL routines, 
which include bias subtraction, flat fielding, order tracing and 
extraction, and wavelength calibration.  
Radial velocities were extracted via multi-order cross-correlations 
using the stellar spectrum (one per CCD) with the highest 
signal-to-noise ratio (S/N) as a template.

 \emph{HARPS} and \emph{HARPS-N} 
are  fibre-fed, cross-dispersed,   \'echelle spectrographs  
 ($R \approx 115\,000$), which are designed to achieve a
very high precision    
and long-term RV measurements.  
We gathered nine spectra with the HARPS spectrograph 
\citep{Mayor03} mounted at the ESO 3.6 m 
telescope of La Silla observatory (Chile), between Oct  2016 and 
Nov 2017, as part of the observing programmes  
098.C-0860, 099.C-0491, and 0100.C-0808.  
We also collected 13 spectra with the HARPS-N spectrograph 
\citep[][]{Cosentino2012} attached at the 
Telescopio Nazionale Galileo (TNG) of Roque de los Muchachos Observatory 
(La Palma, Spain), between Oct 2016 and Jan 2018, during the 
observing programmes CAT16B\_61, CAT17A\_91, A36TAC\_12, and OPT17B\_59. 
We reduced the data using the dedicated off-line HARPS and HARPS-N 
pipeline and extracted the RVs via cross-correlation with a K5 numerical 
mask \citep{1996A&AS..119..373B, 2002A&A...388..632P}. 
The pipeline also provides the bisector inverse slope (BIS) and FWHM of the cross-correlation function (CCF), and the 
log\,R$^\prime_\mathrm{HK}$ activity index of the \ion{Ca}{ii}~H~\&~K 
lines; these are all listed in Table~\ref{Table: RV measurements}.  
Since the   pipelines do not derive the uncertainties for BIS and FWHM we have
   assumed error bars twice as large as the corresponding RV uncertainties in our analysis. 
 The  spectra  have S/N per pixel at 5500\,\AA~in the range 20 -- 45, except for   
one  of the HARPS    and two of the HARPS-N 
measurements with \mbox{S/N < 20} that were not used in the RV analysis.  
In the sixth column the RVs used in the Gaussian Process (GP) regression analysis (Sect.~\ref{Subsection: Gaussian Process}) are
denoted, and in the seventh column we list  
the division of chunks used in the floating chunk offset (FCO)
technique in Sect.~\ref{Subsect: Floating chunk offset method}.


\section{Stellar analysis} \label{Section: Stellar analysis}
  
 The stellar mass and radius needed for the transit and RV analyses can be determined in a number of ways.  
 In this paper we have used   several different methods which requires stellar fundamental parameters
 as input (\teff, \feh, \logg, \rhostar, and distance).

\subsection{Spectral analysis} \label{Subsection: spectral analysis}

In order to derive the stellar fundamental parameters \teff, \logg, and \feh, we analyzed 
the co-added HARPS-N \mbox{(S/N = 89)} and HARPS \mbox{(S/N = 94)} spectra   with the spectral analysis package Spectroscopy Made Easy 
\citep[{\tt{SME}};][]{vp96,vf05,pv2017}. Utilising grids of stellar atmosphere models, based on pre-calculated 1D or 3D, 
local thermal equilibrium (LTE) or non-LTE models,  {\tt{SME}}  calculates, for a set of given stellar parameters, synthetic spectra of 
stars and fits them to observed spectra using a $\chi^2$-minimising procedure. We used the non-LTE  {\tt{SME}}  
version 5.2.2 and the \texttt{ATLAS 12} model spectra \citep{Kurucz2013}  to fit 
spectral features sensitive to different photospheric parameters. We followed the procedure in 
\citet{2017A&A...604A..16F}. In summary, we used  
the profile of the line wings of the 
H$_\mathrm{\alpha}$ and H$_\mathrm{\beta}$ lines  to determine the effective temperature, 
\teff \citep{fuhrmann93, fuhrmann94}. The line cores were excluded owing to its origin in layers 
above the photosphere. The stellar surface gravity, \logg, was estimated from the  line wings of the 
Ca\,{\sc i}~$\lambda \lambda$6102,  6122, 6162 triplet, and the Ca\,{\sc i} $\lambda$6439 line. 
The Mg\,{\sc i} $\lambda \lambda$5167,  5172,  5183 triplet,  which also
can be used to determine \logg, was not used because of problems with the 
density of metal lines contaminating the shape of the wings of the Mg lines. 
The microturbulent velocity, \vmic, and the macroturbulent velocity, \vmac, 
were fixed to 0.5 and 1~\kms, respectively \citep{Doyle2014, 2015A&A...584L...2G}.
The projected stellar rotational velocity, \vsini,  
  and the 
metal abundances 
[Fe/H] and [Ca/H] (needed for the \logg~modelling)  were estimated by fitting the profile of several clean and unblended 
metal lines between 6\,100 and 6\,500~\AA. 
The model was also in agreement with Na doublet $\lambda \lambda$5889 and  5896, 
which showed no signs of interstellar absorption.  
 The results   are listed in  Table~\ref{Table: Spectroscopic parameters}. 
We note that the spectral type of the star is at the lower end for accurate modelling with {\tt{SME}}
due to the weak line wings of the hydrogen and calcium lines, large amount of metal lines interfering with the line profiles,    
low S/N due to the faintness of the star, and  uncertainties
of model atmospheres of cool stars below $\sim 4500$~K.   
 
In  addition to  modelling, we therefore also   used the {\tt{SpecMatch-Emp}}  code \citep{2017ApJ...836...77Y}. 
This code is an algorithm for characterising the properties of stars based on their optical spectra. 
The observed spectra are compared to   a dense spectral library of  
404 well-characterised stars (M5 to F1) observed by Keck/HIRES with high resolution ($R \sim 55\,000$)
and high S/N ($>100$). 
Since the code relies on empirical spectra it performs particularly well for stars $\sim$K4 and later,
which are difficult to model with spectral synthesis models such as {\tt{SME}}. 
However, in extreme cases, such as extremely metal poor or rich stars,
the code could fail since the library includes very few  such stars in each temperature bin. The 
{\tt{SpecMatch-Emp}} code directly yields stellar radius rather than the surface gravity since the library stars 
typically have their radii calibrated using interferometry and other techniques. 
The direct output is  thus  \teff, \rstar, and \feh. 
We note that since the HARPS data suffers from a wavelength gap around 5320 \AA~because the spectrum is recorded on two separate CCD chips, 
the  HARPS-N results should be more accurate.  
Following \citet{2018AJHirano}, prior to the analysis 
we converted  the co-added HARPS and HARPS-N spectra into the  
format of Keck/HIRES spectra that is used by {\tt{SpecMatch-Emp}}. 
In doing so, we made certain that the edges of neighbouring 
\'echelle orders overlapped in wavelength. For the HARPS spectra, the gap 
region was replaced with
a slowly varying polynomial function where each flux relative error is 100\%. The 
validity of analysing spectroscopic data from HARPS, HARPS-N, NOT/FIES, and Subaru/HDS 
with SpecMatch-Emp has been tested by \citet{2018AJHirano}. 
The {\tt{SpecMatch-Emp}}  results and
literature values agree with each other for \teff~and stellar radii mostly within $1\sigma$. 
The \feh~values   sometimes show a moderate disagreement, but are basically consistent within
$2\sigma$.  
The results  are listed in  Table~\ref{Table: Spectroscopic parameters} and  \ref{Table: Comparison stellar parameters}. 

The effective temperatures derived with {\tt{SME}}  and {\tt{SpecMatch-Emp}} HARPS-N are in excellent agreement. 
The metallicities are in agreement within $1\sigma$.  
Since the results are in such   good agreement, and since we have no clear motivation of preferring one model over the other  
despite
their respective possible issues, we adopted an average of the modelled effective temperatures and metallicities from  {\tt{SME}} and {\tt{SpecMatch-Emp}} HARPS-N. 
Our adopted \teff~is also consistent with the findings of \citet[][$4\,591\pm50$~K]{2018AJ....155..136M}, whereas  
their metallicity ($-0.18\pm0.08$~dex) is  lower than our average value, but is   within  $2\sigma$.  
For \logg~we adopted the value from {\tt{SpecMatch-Emp}} HARPS-N coupled with the 
\citet{Torres2010} calibration equations  
(see Sect.~\ref{Subsection: stellar mass and radius}) owing to  difficulties in modelling
the Ca lines accurately with  {\tt{SME}}  for this type of star.   
The \logg~from {\tt{SpecMatch-Emp}} is also in perfect
agreement with  the adopted stellar mass and radius (Table~\ref{Table: Final stellar parameters}) and
with the results from  {\tt{PARAM\,1.3}} (Sect.~\ref{Subsection: stellar mass and radius}).  
It is in addition in excellent agreement with the results from  \citet[][log\,{\it g$_\star = 4.59\pm0.10$}]{2018AJ....155..136M}.
Within $1\sigma$, our resulting \teff~and \feh~ are also in agreement 
 with the listed parameters in the Ecliptic Plane Input Catalog  \citep[EPIC; ][]{2016ApJS..224....2H},  
\teff~$= 4\,653\pm95$~K and \feh~$= -0.02\pm0.2$ (dex). However, we find that the listed \mbox{\logg~$= 2.76\pm0.43$ (cgs)},  
\mbox{\rstar~$= 6.9\pm4.7$~\Rsun},  and  the stellar density of $3\times 10^{-3}$~\gc, which 
points to an evolved giant star at a distance of $1\,159\pm555$~pc, are erroneous and 
in major disagreement with our spectral  analysis, 
the Gaia distance by a factor 
of ten (Sect.~\ref{Subsection: stellar mass and radius}), and  the stellar density derived  
by our transit modelling (Sect.~\ref{Section: transit analysis}).   
For the  projected rotational velocity, \vsini, we adopted the value determined with {\tt{SME}}. 
  
Using the 
\citet{Straizys1981} calibration scale for dwarf stars,  the 
spectral type is defined as \stype. 
The adopted stellar  parameters are listed in Table~\ref{Table: Final stellar parameters}.

\begin{table}
 \centering
 \caption{Stellar mass and radius of \target~as derived from different methods.
 Typical values   for a  \stype~star are listed as comparison. 
 }
\begin{tabular}{lcc}
 \hline\hline
     \noalign{\smallskip}
~Method~~~~~~~~~~~~~~~~~~~~~~~~~~~~~~~  & \mstar~~~~~~~~~~~~~~~ &       ~\rstar ~~~~~~~~~~~~    \\
~~~~~~~~~~~~~~~~~~~~~~~~~~~ &  (\Msun)~~~~~~~~~~~~~ & (\Rsun)~~~~~~~~~~      \\
    \noalign{\smallskip}
     \hline
\noalign{\smallskip} 
\end{tabular}
\begin{tabular}{lll}
Gaia$\tablefootmark{a}$ 
& \ldots    & \sradiusgaia  \\

 {\tt {SpecMatch-Emp}}/Torres    
 & \smasssp$\tablefootmark{b}$ & \sradius$\tablefootmark{c}$  \\ 

  {\tt {PARAM 1.3}}      
 & \smassp    & \sradiusp  \\

   
  {\tt {BASTA}}          
 & \smassb    & \sradiusb  \\
 
   Spectral\,type \stype         
 & \smassspectral & \sradiusspectral      \\ 

      \noalign{\smallskip} \noalign{\smallskip}
\hline 
\end{tabular}
\tablefoot{
\tablefoottext{a}{Radius calculated from Gaia parallax, our modelled \teff, and apparent visual magnitude.}
\tablefoottext{b}{Coupling the {\tt{SpecMatch-Emp}} \mbox{HARPS-N} modelling  with  the calibration equations from  \citet{Torres2010}.}
\tablefoottext{c}{Direct result from {\tt{SpecMatch-Emp}}.}
}
 \label{Table: Comparison stellar parameters}
\end{table}

\begin{table}
 \centering
 \caption{Adopted  stellar parameters of \target.}
\begin{tabular}{ll}
 \hline\hline
     \noalign{\smallskip}
 Parameter &\target~    \\
    \noalign{\smallskip}
     \hline
\noalign{\smallskip} 

Effective temperature \teff$\tablefootmark{a}$~(K)\dotfill &\stempave \\ \noalign{\smallskip}
Surface gravity $\log(g_\star)\tablefootmark{b}$   (cgs)        \dotfill & \sloggsp \\  \noalign{\smallskip}              
Density  $\rho_\star\tablefootmark{c}$   (g~cm$^{-3}$)\dotfill & \denspyaneti  \\ \noalign{\smallskip}
Metallicity [Fe/H]$\tablefootmark{a}$ (dex)   \dotfill & \sfehave \\ \noalign{\smallskip}
Rotational velocity \vsini$\tablefootmark{d}$~(\kms)  \dotfill&  \svsini        \\  \noalign{\smallskip}

Mass $M_\star\tablefootmark{e}$  (\Msun)  \dotfill &  \smassmean\\ \noalign{\smallskip}
Radius $R_\star\tablefootmark{f}$   (\Rsun)  \dotfill & \sradiusgaia \\ \noalign{\smallskip}
Luminosity $L_\star\tablefootmark{f}$    (\Lsun)\dotfill & \lstargaia    \\ \noalign{\smallskip}

Spectral type   \dotfill&  \stype  \\ \noalign{\smallskip}

Rotation period  ~(days) \dotfill&  \prot   \\ 

    \noalign{\smallskip} \noalign{\smallskip}
\hline 
\end{tabular}
\tablefoot{
\tablefoottext{a}{Average  from  {\tt{SME}}   HARPS and HARPS-N, and {\tt{SpecMatch-Emp}} HARPS-N.}
\tablefoottext{b}{{\tt{SpecMatch-Emp}} HARPS-N.}
\tablefoottext{c}{Derived from transit modelling.} 
\tablefoottext{d}{Average  from  {\tt{SME}}   HARPS and HARPS-N.}
\tablefoottext{e}{{\tt{SpecMatch-Emp}}/Torres and \texttt{BASTA}.}
\tablefoottext{f}{Calculation based on the Gaia parallax, our modelled \teff, and apparent visual magnitude.}
}
 \label{Table: Final stellar parameters}
\end{table}

%

\begin{figure*}
\centering
  \begin{subfigure}[b]{0.4\textwidth}
    \includegraphics[width=\textwidth]{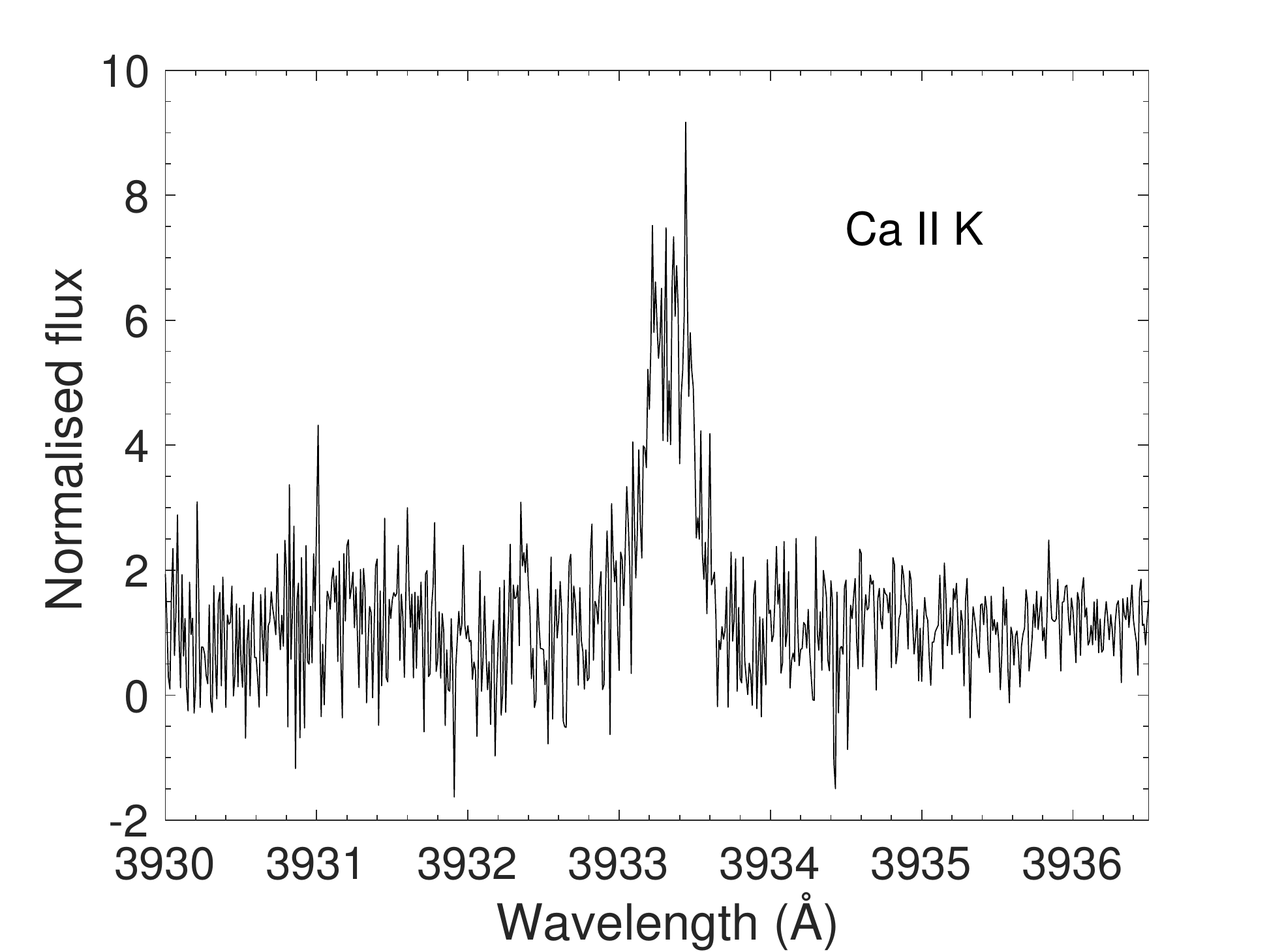}
    \label{subfigure: CaII K}
  \end{subfigure}
  \begin{subfigure}[b]{0.4\textwidth}
    \includegraphics[width=\textwidth]{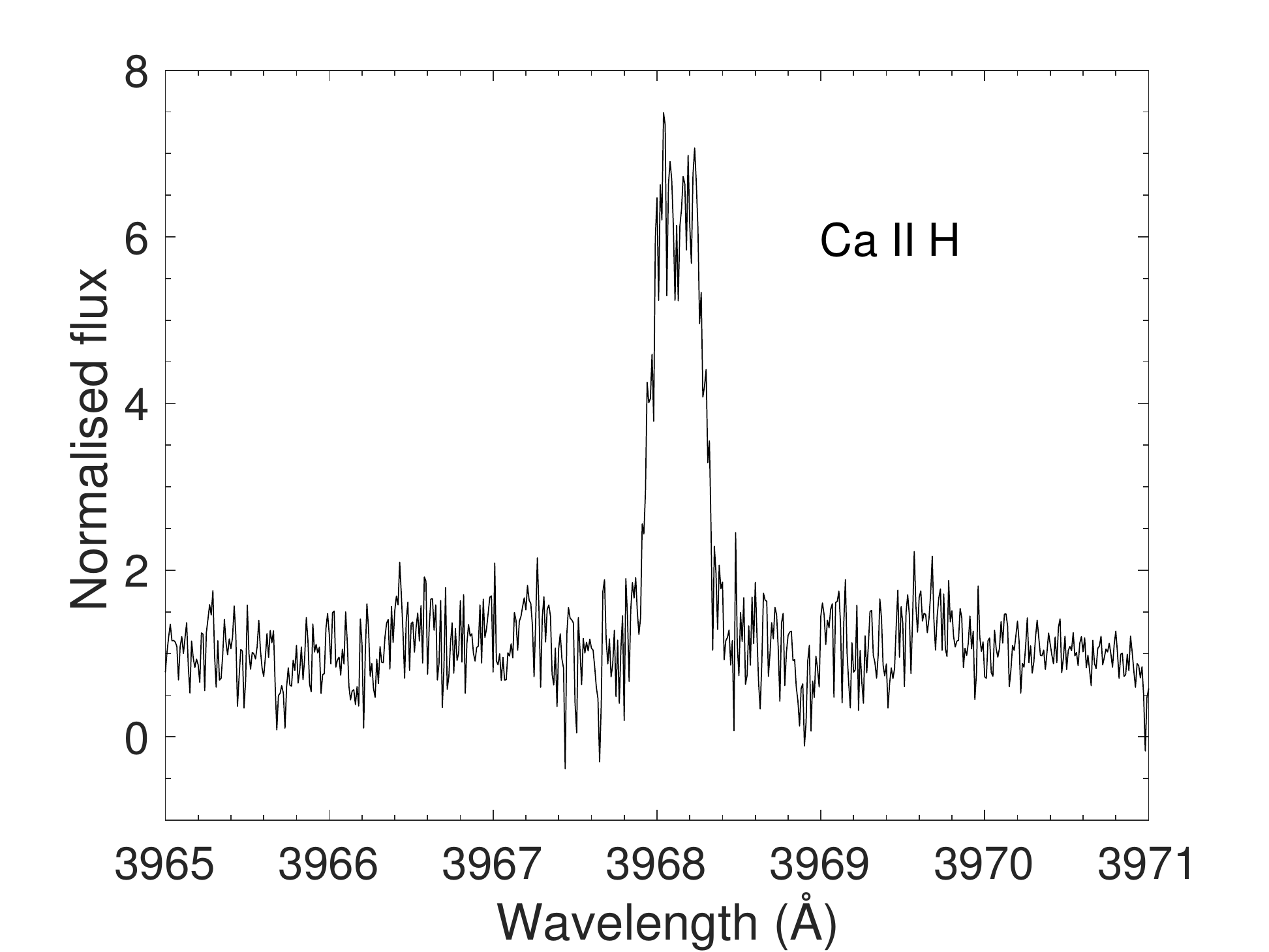}
    \label{subfigure: CaII H}
  \end{subfigure}
   \caption{Cores of the  \ion{Ca}{II} K \emph{(left)} and H \emph{(right)} lines are seen  in emission indicating activity of \target. }
     \label{Figure: CaII K and H}
\end{figure*}


\subsection{Stellar mass and radius} \label{Subsection: stellar mass and radius}
We calculated the stellar radius by combining the distance obtained from  the 
Gaia DR2\footnote{\url{http://gea.esac.esa.int/archive/}}  parallax (\parallaxgaia~mas corresponding
to a distance of $115.8\pm0.7$~pc) with 
our spectroscopically derived \teff~and the apparent visual magnitude. 
We added 0.1~mas in quadrature to the parallax uncertainty to account for systematic errors 
of Gaia's astrometry \citep{2018arXiv180409376L}.  
We first calculated the luminosity from  
the relations 
\mbox{$M_\mathrm{V\star} =  V - 5\times \log_{10}(\mathrm{d}) + 5 - A_\mathrm{V}$},  
 \mbox{$M_\mathrm{bol\star} = M_\mathrm{V\star} + \mathrm{BCv}$}, 
 and  \mbox{$M_\mathrm{bol\star} - M_\mathrm{bol, \odot} = -2.5 \times \log_{10}(L_\star/L_\odot)$}, 
  where
$M_\mathrm{bol}$ is the absolute bolometric magnitude,  
BCv is the temperature-only dependent bolometric correction   of \BCv~\citep{2000asqu.book.....C},   
 $A_\mathrm{V}$ is the visual extinction here assumed to be zero given the proximity of \target, 
 and $M_\mathrm{bol,\odot}=+4.74$.  
The stellar luminosity was found to be  \mbox{\lstargaia~\Lsun}. This value was then used to calculate the
stellar radius with    
\mbox{$L_\star =  4\, \pi\,R_\star^2\, \sigma\, T_\mathrm{eff}^4$},  
which was  found to be \mbox{\sradiusgaia~\Rsun} (Table~\ref{Table: Comparison stellar parameters}). 
This value is in excellent agreement    with  
the spectroscopic radius derived using  {\tt{SpecMatch-Emp}} (\sradius~\Rsun), but larger than 
that found by  \citet[][$0.67\pm0.02$~R$_\odot$]{2018AJ....155..136M} although still within  $1\sigma$.  
An extinction close to zero is also  supported by  
the absence of interstellar components in the \ion{Na}{I} doublet at 5889~\AA~and   by
following 
the method outlined in \citet{gandolfi08}. This method   
 adopted the extinction law by \citet{Cardelli1989} and assumed    \mbox{$R_\mathrm{V} = A_\mathrm{V}/ E_\mathrm{B-V} = 3.1$}.  
A spectral energy distribution  was then fitted 
using synthetic colours calculated ad hoc from the  {\tt{BT-NEXTGEN}} low resolution spectrum
model \citep{Allard2011} with the \target~parameters.

The stellar mass must be modelled and this is carried out with four different methods. These models also produce a 
stellar radius, which is, however, only used as a comparison with the radius derived above. 
Coupling   the  {\tt{SpecMatch-Emp}} modelling with the \citet{Torres2010} calibration equations,  we  
find a stellar mass and surface gravity  of \mbox{\smasssp~\Msun} and  \mbox{\logg = \sloggsp~(cgs)}, respectively. 
The Torres equations were calibrated with 95 eclipsing binaries where the masses and
radii were known to better than 3~\%.
The \logg~is in agreement with the {\tt{PARAM\,1.3}} result below and with \citet{2018AJ....155..136M},  
but higher than obtained with {\tt{SME}}, although
still within   $1\sigma$. The corresponding stellar density is in   agreement   with
the density found from the transit modelling in Sect.~\ref{Section: transit analysis}.  
 
We have also used the Bayesian 
{\tt{PARAM\,1.3}}\footnote{\url{http://stev.oapd.inaf.it/cgi-bin/param_1.3}}  \citep{daSilva2006} on-line applet   to 
obtain mass,  radius, and age  using the  {\tt{PARSEC}} isochrones from \citet{2012MNRAS.427..127B}.  
The required input is   parallax,  \teff, \feh, and apparent visual magnitude. 
The results were a stellar mass of  \mbox{\smassp~\Msun}, 
 radius of  \mbox{\sradiusp~\Rsun},  
surface gravity of \sloggp (cgs), and age of \age~Gyrs.  
The mass and radius were also   estimated with 
 the \citet{2011MNRAS.417.2166S} calibration equations built on the basis of 90 detached eclipsing binaries 
 with masses up to 3~\Msun.  The    input parameters are the stellar density
(derived from transit modelling in Sect.~\ref{Section: transit analysis}),   together with the spectroscopically derived \teff~and \feh.  
Since the derived density  has large uncertainties, this is propagated to the   
 mass (\mbox{\smasssouth~\Msun}) and radius
  (\mbox{\sradiussouth~\Rsun}). This calibration method is thus not very useful for modelling   \target, and
   is therefore 
 not used  for the mass estimate.  
Finally, we    used the BAyesian STellar Algorithm 
\citep[{\tt{BASTA}};][]{Silvaaguirre2015}.\ The \texttt{BASTA} model uses a 
Bayesian approach to isochrone grid-modelling and fits observables to a 
grid of BaSTi isochrones  \citep{2004ApJ...612..168P}. We fit 
spectroscopic ($T_{\text{eff}}$, $\log g_\star$, $\text{[Fe/H]}$) and 
photometric ($\rho_\star$) constraints and find  a stellar mass and radius of 
\smassb~M$_{\odot}$ and \sradiusb~R$_{\odot}$, respectively, 
and an age of \agebasta~Gyr.

All  estimates of the stellar mass   are in very good agreement and are listed in  Table~\ref{Table: Comparison stellar parameters}, 
 along with a typical mass and radius for a \stype~star for comparison. 
We choose to  adopt a value of \smassmean~\Msun~since  two of the models give this  stellar mass, and 
the third ({\tt{PARAM\,1.3}}) only slightly higher.  
This  mass is also in excellent agreement with    \citet[][\mbox{$M_\star = 0.70\pm0.02$~\Msun}]{2018AJ....155..136M}. 
 
All  adopted stellar parameters are listed in Table~\ref{Table: Final stellar parameters}.

\subsection{Stellar activity and rotation period} \label{Section: stellar activity}

Before analysing the RV measurements, 
we need to check whether they are affected by stellar activity. 
 Photometric variability in solar-like stars can be caused by stellar 
activity, such as spots and plages, on a timescale comparable to the 
rotation period of the star. The presence of active regions coupled to 
stellar rotation distorts the spectral line profile, inducing periodic 
and quasi-periodic apparent RV variation, which is commonly referred to as 
RV jitter.

The presence of active regions hampers our capability of detecting small 
planets using the RV method. This is because the expected RV wobble 
induced by small planets is of the same order of magnitude, or   even 
smaller, than the activity-induced jitter. 
Nevertheless, if the orbital period of a planet is much smaller than the stellar 
rotation period, then the correlated noise due to stellar rotation can 
easily be distinguished from the planet-induced RV signal 
\citep{Hatzes11}. An inspection of the \emph{K2} light curve shows, 
quasi-periodic photometric variations with a typical peak-to-peak amplitude of 
about $\sim0.4 - 0.5$~\%. Given the
spectral type of the star, the variability is very likely associated with 
the presence of spots on the photosphere of the star, combined with 
stellar rotation and/or its harmonics\footnote{The presence of active 
regions at different longitudes can induce photometric signals at 
rotation period harmonics.}. The light curve also shows that spots evolve 
with a
timescale that is comparable to the duration of the \emph{K2} observations 
(about \mbox{80~days}).

Inspecting the \ion{Ca}{II} H $\&$ K lines in the HARPS-N spectra, we 
find that both lines are seen in emission as shown in \mbox{Fig.~\ref{Figure: CaII K and H}}. 
We measure an average \ion{Ca}{II} 
chromospheric activity index, \lgr~in Table~\ref{Table: RV measurements} of 
$-4.668\pm 0.059$ and $-4.658\pm 0.069$ from HARPS and HARPS-N, respectively,  
 indicating that the star is 
moderately active.  

Using the code 
SOAP2.0\footnote{\url{http://www.astro.up.pt/resources/soap2/}} \citep{2014ApJ...796..132D} and 
adopting the stellar parameters reported in 
Table~\ref{Table: Final stellar parameters}, 
an average peak-to-peak variation of 0.45~\%, the same limb-darkening coefficients  (LDCs) used in the transit
modelling in Sect.~\ref{Section: transit analysis}, and modelling two starspots 
with a size relative to the star of 0.07, we  found that the 
expected RV jitter is  $\sim4$~m~s$^{-1}$.

The upper panel of Fig.~\ref{Figure: Davide periodograms} (appendix~\ref{Sect: appendix figures})
represents the generalised Lomb-Scargle 
\citep[GLS;][]{2009A&A...496..577Z} periodogram of the \emph{K2}  light curve of \target. 
Prior to computing the periodogram, we removed the transit signals using 
the best-fitting transit model derived in Sect.~\ref{Section: transit analysis}.  
We also subtracted a 
linear fit to the \emph{K2}  data to remove the flux drift often observed across 
many \emph{K2}  stars, which is likely caused by slow changes in the spacecraft 
orientation and/or temperature. The remaining panels of Fig.~\ref{Figure: Davide periodograms} show the 
GLS periodograms of the RV, BIS, FWHM, and \lgr~extracted from the 
FIES, HARPS, and HARPS-N data, which were   combined after  
subtracting the corresponding means of the data sets of each instrument (Table~\ref{Table: RV measurements}). The 
FAPs were determined following the bootstrap 
technique described in \citet{1997A&A...320..831K}.

The periodogram of the \emph{K2}  light curve displays a very significant peak 
(FAP $\ll 1\%$) at $30\pm5$ days (vertical dashed blue line in 
 Fig.~\ref{Figure: Davide periodograms}), which we interpreted as being the rotation period of the star 
($P_\mathrm{rot}$). Assuming that the star is seen equator-on, this 
value is within the limits obtained from the stellar radius and the 
spectroscopically derived, projected rotational velocity \vsini.  We
found that $P_\mathrm{rot} = 2 \pi R_\star / V$ should be between 9 and 
32~days, including the uncertainties on \vsini~and \rstar.

The dashed vertical red line in Fig.~\ref{Figure: Davide periodograms}  indicates the orbital frequency of 
the transiting planet, whereas the horizontal lines represent the 1\,\% 
FAP. The periodogram of the RV measurements 
displays a peak at the orbital frequency of the transiting planet with a 
FAP of 1\ \%, which has no counterparts in the periodograms of the 
activity indicators, suggesting that this signal is induced by the 
transiting planet. We note  the presence of peaks in the periodograms of  
BIS and FWHM whose frequencies are close to the rotation frequency of 
the star.

\section{Transit modelling} \label{Section: transit analysis}

 We used the orbital period, mid-transit time, transit depth, and transit duration identified by  {\tt{EXOTRANS}}
as input values for more detailed transit modelling with the publically available 
software {\tt{pyaneti}}\footnote{\url{https://github.com/oscaribv/pyaneti}} \citep{2017ascl.soft07003B}, which is also used
in for example \citet{Barragan2016}, \citet{2017AJ....154..123G}, and \citet{2017A&A...604A..16F}.  
{\tt{Pyaneti}} is a  \texttt{PYTHON/FORTRAN} software that infers planet parameters
using Markov chain Monte Carlo (MCMC) methods based on Bayesian analysis.

The {\tt{Pyaneti}} software allows a joint modelling of the transit and RV data.
Stellar activity can, however, only be modelled in  {\tt{pyaneti}} as a coherent signal, not changing in time or phase, which is
only possible
 when the RV observational season is small compared to the evolution timescale of active regions
  \citep[e.g.][]{2018A&A...612A..95B}. Since this  is not the case
 for \target~where the observations extend over 440~days, we
 only used {\tt{pyaneti}} to model the transit data. 
   
 In order  to prepare the light curve for {\tt{pyaneti}} and  reduce the amplitude of any long-term
systematic or instrumental flux variations, we used the  {\tt{exotrending}} \citep{2017ascl.soft06001B} code
to detrend the Vanderburg  transit light curve (Fig.~\ref{Figure: full lightcurve}) by 
fitting a second order polynomial  to the out-of-transit data.
Input to the code is  the mid-time of first transit, $T_0$, and orbital period, $P_\mathrm{orb}$.
Three hours around each of the 36 transits
was masked to ensure that no in-transit data were used in the detrending process.

We followed the procedure in \citet{Barragan2016} for the {\tt{pyaneti}}  transit modelling.
For the mid-time of first transit ($T_0$), the orbital period, $P_\mathrm{orb}$, the scaled 
orbital distance ($a/R_\star$), 
the planet-to-star radius  ratio (\rplanet/\rstar),
and the impact parameter \mbox{($b \equiv a \cos (i)/R_\star$)}, we set uniform priors meaning that we adopted rectangular distributions over
given ranges of the parameter spaces. 
The value $T_0$ is measured relatively precise compared to the cadence of the light curve, and
$P$ is measured very precise because of the large number of transits (36), and the absence of measurable transit timing
variations. 
The ranges are thus \mbox{$T_0 = [7394.03887, 7394.05887]$ (\bjdtdb - 2450000)}~days,
$P= [2.17249, 2.17649]$~days,  \mbox{$a/R_\star= [1.1, 50]$}, \mbox{$b= [0, 1]$}, and \mbox{\rplanet/\rstar~$= [0, 0.1]$}. 
Circular orbit was assumed, hence
the eccentricity ($e$) was fixed to zero, and  the argument of periastron,  $\omega$, was set to $90^\circ$.
The transit models were integrated over ten steps to
account for the long integration time (29.4~minutes) of \emph{K2} \citep{Kipping2010}.
We adopted the quadratic limb darkening equation by \citet{Mandel2002}, which uses the linear 
and quadratic  coefficients $u_1$ and $u_2$, respectively. We  followed the parametrisation 
$q_1 = (u_1 + u_2)^2$ and $q_2 = 0.5 u_1(u_1+u_2)^{-1}$ from \citet{Kipping2013}. 
We first ran a fit using uniform priors for the LDCs and noticed that the 
LDCs were not well constrained by the light curve.
This is because the  LDCs are not well constrained for small planets using uniform priors 
\citep[e.g.][]{2013A&A...549A...9C}.  
Thus, we used the on-line applet\footnote{{\url{http://astroutils.astronomy.ohio-state.edu/exofast/limbdark.shtml}}}  
written by
\citet{Eastman2013} to interpolate the \citet{Claret2011} limb 
darkening tables to the spectroscopic parameters of \target~to estimate $u_1$ and $u_2$. These  
values were used to set Gaussian priors to $q_1$ and $q_2$ LDCs with 0.1 error bars. The planetary and orbital 
parameters were consistent for both LDC prior selections. We used the model with Gaussian priors on 
LDC for the final parameter estimation.

\begin{figure}[!t]
 \centering
  \resizebox{\hsize}{!}
   {\includegraphics{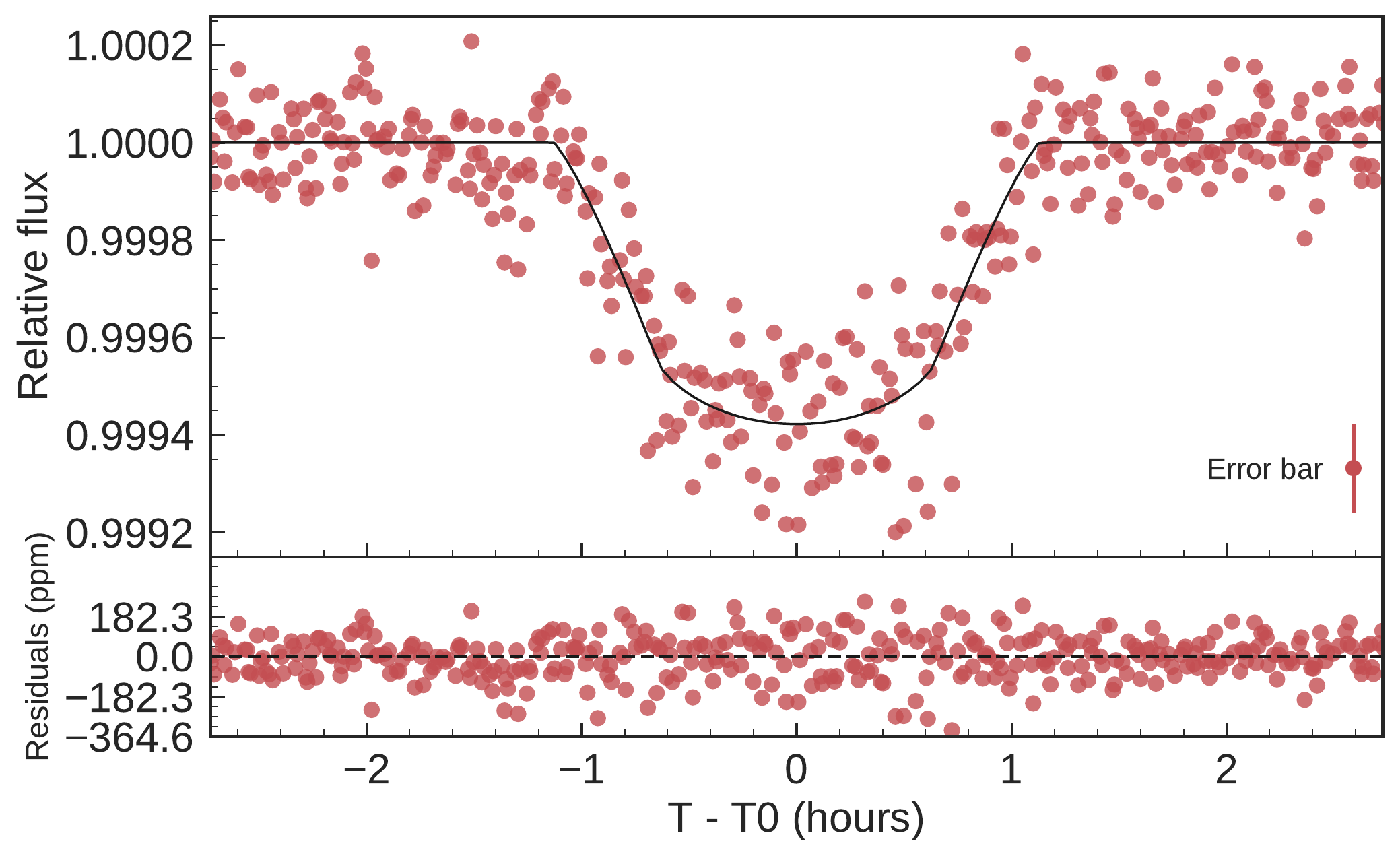}}
   \caption{Transit light curve folded to the orbital period of \target~and residuals. The red points 
   indicate the \emph{K2} photometric data.  
   The solid line indicates the {\tt{pyaneti}} best-fitting transit model.}
      \label{Figure: pyaneti folded lightcurve and model}
 \end{figure}

We explored the parameter space with 500 independent chains created randomly inside the prior ranges. 
We checked for convergence each 5\,000 iterations. Once convergence was found, we used the last 5\,000 
iterations with a thin factor of 10 to create the posterior distributions for the fitted parameters. 
This led to a posterior distribution of 250\,000 independent points for each parameter. 
The posterior distributions for all parameters were smooth and unimodal. 
The final planet parameters are listed in Table~\ref{Table: Orbital and planetary parameters}, and 
the resulting    stellar density is listed in Table~\ref{Table: Final stellar parameters}. 
The folded light curve and best-fitted model (binned to the \emph{K2} integration time to allow comparison with the data) 
is shown in Fig.~\ref{Figure: pyaneti folded lightcurve and model}.

\section{Radial velocity modelling} \label{RV analysis}

\subsection{Gaussian process regression}\label{Subsection: Gaussian Process}
We used a GP  regression model described by \citet{2017AJ....154..226D} 
to simultaneously model the planetary signal and correlated noise associated with stellar activity. This code 
is able to  fit a non-coherent signal, assuming that activity acts as a signal whose period is
 given by the rotation period of the star, and whose amplitude and phase change on a time
scale given by the spot evolution timescale.
Gaussian process describes a stochastic process as a covariance 
matrix whose elements are generated by user-specified kernel functions. With suitable choice of the 
kernel functions and the hyperparameters that specify these functions,  GP can be used to model a wide range 
of stochastic processes. GP regression has been successfully applied to the RV analysis 
of several exoplanetary systems where correlated stellar noises cannot be ignored, for example CoRoT-7 \citep{2014MNRAS.443.2517H}, 
Kepler-78 \citep{2015ApJ...808..127G},  and Kepler-21 \citep{2016AJ....152..204L}.

Magnetic active regions on the host star, coupled with stellar rotation, result in quasi-periodic 
variations in both the measured RV and flux variation. 
Given their similar physical origin, both the quasi-periodic 
flux variation and correlated stellar noise in the RV measurements encode physical information about the host stars, for example the stellar rotation period and lifetime of starspots. This information is reflected in the 
hyperparameters of GP used to model these effects. In particular, there is a good correspondence between the stellar 
rotation period and the period of the covariance, $T$, while the correlation timescale, $\tau$, and weighting 
parameter, $\Gamma$, together determine the lifetime of starspots. 
We can thus model both the rotational modulation in the light curve and the correlated noise in RV as GPs.

Since the \emph{K2} light curve was measured with 
a high-precision, high temporal sampling, and an almost continuous temporal coverage, we trained our GP model on the \emph{K2} light curve  after removal of the transits. The constraints on 
the hyperparameters were then used as priors in the 
subsequent RV analysis. We used the covariance matrix and the likelihood function described by \citet{2017AJ....154..226D}  
and adopted a quasi-periodic kernel,  
\begin{equation}
\label{covar}
C_{i,j} = h^2 \exp{\left[-\frac{(t_i-t_j)^2}{2\tau^2}-\Gamma \sin^2{\frac{\pi(t_i-t_j)}{T}}\right]}+\left[\sigma_i^2+\sigma_{\text{jit}}(t_i)^2\right]\delta_{i,j}
,\end{equation}
where $C_{i,j}$ is an element of the covariance matrix, and $\delta_{i,j}$ is the Kronecker delta function. 
The hyperparameters of the kernel are the covariance amplitude $h$, $T$,  $\tau$,  the time of $i$th observation, $t_i$,     
and  $\Gamma$  which
 quantifies the relative importance between the squared exponential and periodic parts of the kernel. 
For the planetary signal,  we assumed a circular Keplerian orbit. The corresponding parameters 
are the RV semi-amplitude, $K$, the orbital period, $P_{\text{orb}}$, and the time of conjunction, $t_{\text{c}}$. Since 
our data set consists of observations from several observatories, we included a separate jitter parameter,  $\sigma_{\text{jit}}$, 
to account for additional stellar/instrumental noise,  
and a systematic offset, $\gamma$, for each of the instruments  
(Table~\ref{Table: RV measurements}). The orbital period and time of conjunction are 
much better measured using the transit light curve. 
We thus imposed Gaussian priors on $P_{\text{orb}}$ and 
$t_{\text{c}}$ as derived from the \emph{K2} transit modelling. 
We imposed a prior on $T$ using the stellar rotation period measured from the periodogram ($30\pm5$~days). 
The scale parameters $h$, $\tau$, $K$, and the jitters were sampled uniformly in log space, basically imposing a
Jeffrey's priors. Uniform priors  were imposed on the systematic
offsets.

The  likelihood function has the following form:
\begin{equation}
\label{likelihood}
\log{\mathcal{L}} =  -\frac{N}{2}\log{2\pi}-\frac{1}{2}\log{|\bf{C}|}-\frac{1}{2}\bf{r}^{\text{T}}\bf{C} ^{-\text{1}} \bf{r}
,\end{equation}
where $\mathcal{L}$ is the likelihood, $N$ is the number of data points, $\bf{C}$ is the covariance matrix, 
and $\bf{r}$ is the residual vector 
(the observed  value minus the model value). 
The model includes the RV variation induced by the planet and a constant offset for each instrument.

\begin{figure}[!t]
 \centering
  \resizebox{\hsize}{!}{
   \includegraphics{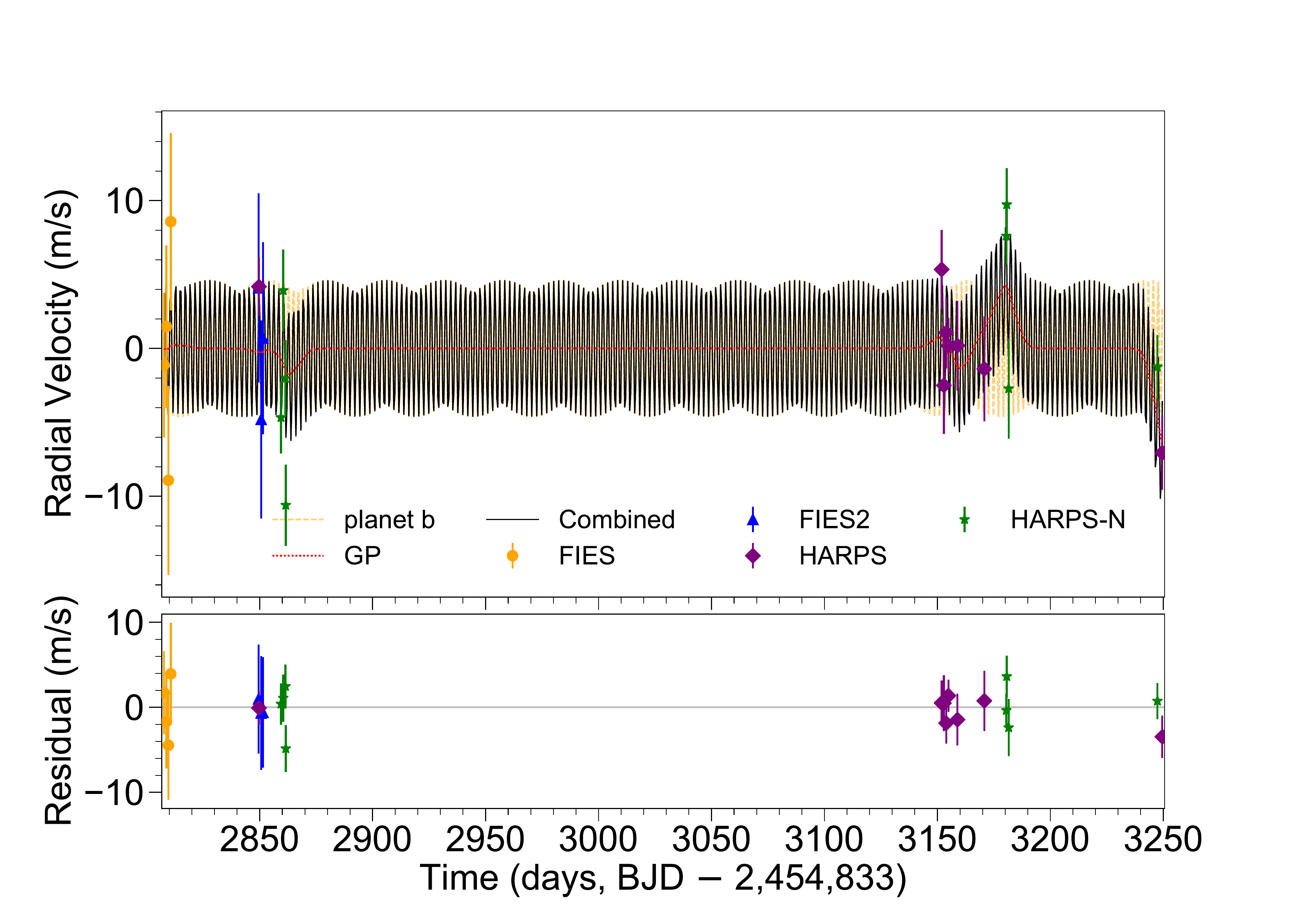}}
   \caption{Measured RV variation  of \target~from HARPS-N (green stars), HARPS (purple diamonds),  
   FIES (yellow circles), and FIES2 (blue triangles). The black solid line is the best fit from the GP regression model of the correlated stellar noise
   and the signal from \target b. The signal from  planet~b is  shown by the yellow dashed line, and the GP
   regression model of correlated stellar noise by the red dotted line. The lower panel shows residuals of the fit.}
      \label{Figure: 1411.pdf}
 \end{figure}
 \begin{figure}[!t]
 \centering
  \resizebox{\hsize}{!}{
   \includegraphics[width=\linewidth]{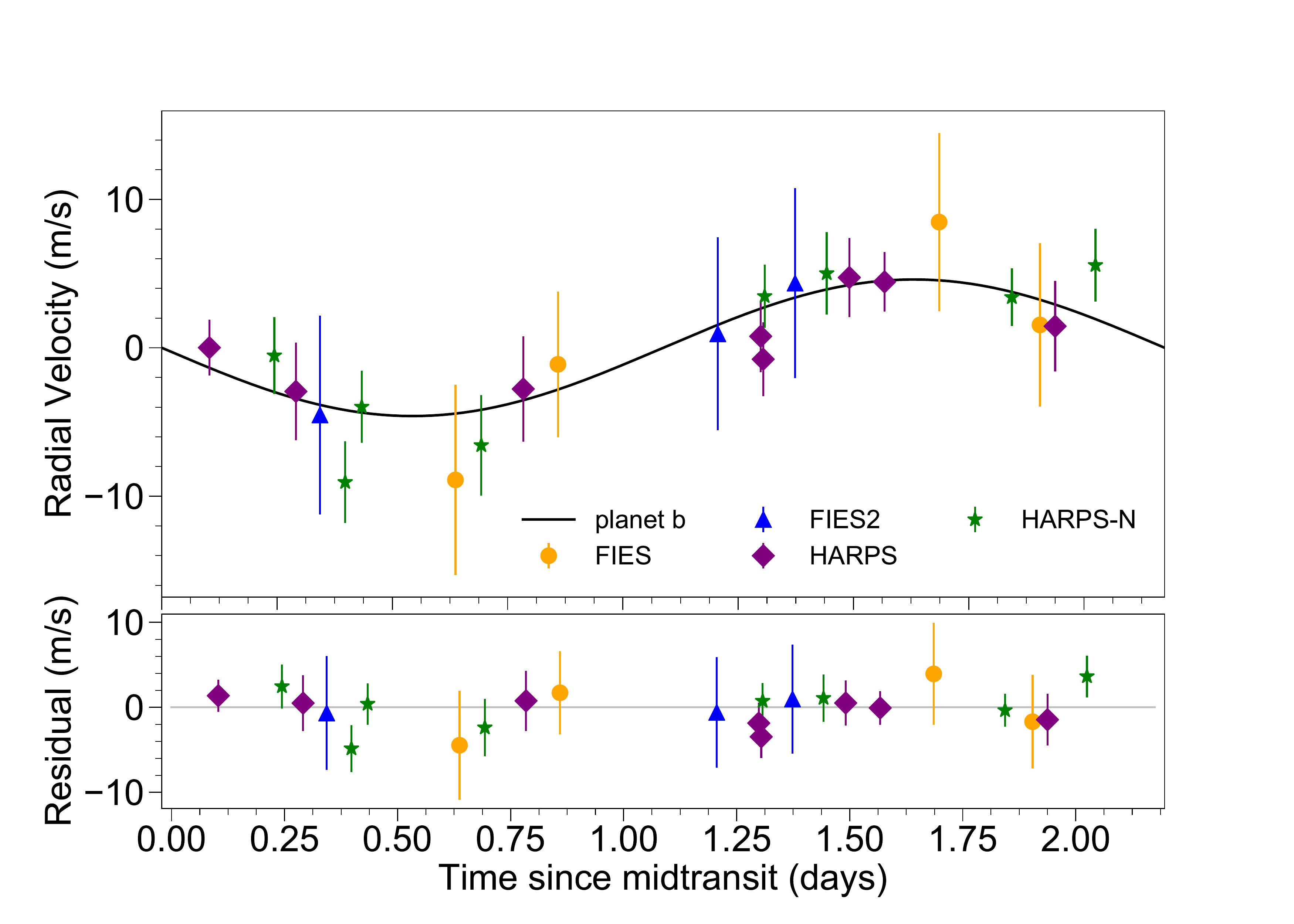}}
   \caption{Radial velocity curve  of \target~phase folded to the orbital period of the planet using the GP 
   regression model.  The data are plotted with the same colour code as in Fig.~\ref{Figure: 1411.pdf}. 
 The resulting \mbox{$K$ amplitude} is \mbox{\kb~\ms}. 
   The lower panel shows residuals of the fit.}
      \label{Figure: 1411b.pdf}
 \end{figure}
%

 
We first located the maximum likelihood solution using the Nelder-Mead algorithm implemented in the {\tt Python} 
package {\tt scipy}. We sampled the posterior distribution using the affine-invariant MCMC implemented in the 
code {\tt emcee} \citep{2013PASP..125..306F}. We started 100 walkers near the maximum likelihood solution.  
We stopped after running the walkers for 5000 links. We checked for convergence by calculating the Gelman-Rubin 
statistics, which dropped below 1.03 indicating adequate convergence. We report the various parameters using 
the median and 16\% -- 84\% percentiles of the posterior distribution. 
The hyperparameters were constrained to 
be \mbox{$\tau$ = $4.8^{+7.3}_{-2.9}$ days}, 
\mbox{$\Gamma$ = $1.28 \pm 0.63$}, and   $T = 28\pm4$~days. 
These were incorporated as priors in the subsequent GP analysis of the RV data.

We followed a similar procedure when analysing the RV data set. We first found the maximum likelihood solution 
and then sampled the posterior distribution with MCMC. 
We removed four isolated 
RV measurements, which were separated by more than approximately two $\tau$ from any neighbouring data points,  from the 
GP modelling. Without neighbouring data points, the stellar variability component 
of these isolated data points are causally disconnected. As a result, GP tends to overfit these data points and 
thus underestimate the planetary signal. The removed RVs are listed in column six, Table~\ref{Table: RV measurements}.  
The RV semi-amplitude for planet \target b was constrained  to 
\mbox{\kb~m~s$^{-1}$}.  
Using the stellar mass derived in Sect.~\ref{Subsection: stellar mass and radius} of \mbox{\smasssp~\Msun}, this translates to a planet 
mass of 
\mbox{\mpbGP~\mearth} with a precision in mass of 30~\%.  
 The amplitude of the correlated stellar noise is 
  \mbox{$h_{\text{rv}} = $ \hGP ~\ms}, which agrees with the SOAP2.0 modelling in
 Sect.~\ref{Section: stellar activity}.  
 The 95\% upper bounds of the jitters were found to be \mbox{$<5.1$~\ms}~(FIES), $<4.7$~\ms~(FIES2), \mbox{$<2.5$~\ms}~(HARPS), and 
 \mbox{$<2.6$~\ms}~(HARPS-N). 
As a comparison, keeping all the RVs with $\mathrm{S/N} > 20$,  we obtain \mbox{$K =$~\kbold~m~s$^{-1}$} corresponding
to a planet mass of  \mbox{\mpbGPold~\mearth}, and    \mbox{$h_{\text{rv}} = $   \hGPold ~\ms}. 
 Figure~\ref{Figure: 1411.pdf}   
 shows the measured RV variation 
 of \target~and the GP model. The folded RV diagram as a function of orbital phase is shown Fig.~\ref{Figure: 1411b.pdf}.
The results are listed in Table~\ref{Table: Orbital and planetary parameters}.

\begin{table*}
\centering
\caption{Final  \target b parameters.}
\begin{tabular}{lcr}
\hline
\hline
\noalign{\smallskip}
Parameter & Units & Value  \\
\noalign{\smallskip}
\hline
\noalign{\smallskip}
              
\multicolumn{3}{l}{\emph{Transit and orbit parameters}}\\
\noalign{\smallskip}                
                ~~~$P_\mathrm{orb}$ &Period (days)\dotfill & \Pb \\
\noalign{\smallskip}                
                ~~~$T_0$ &Time of transit (BJD - 2450000) \dotfill & \Tzerob             \\
\noalign{\smallskip}                          
                ~~~$T_{14}$ &Total duration (hours)\dotfill & \ttotb     \\
\noalign{\smallskip}                                
                ~~~$b$  &Impact parameter\dotfill & \bb  \\
\noalign{\smallskip}
                ~~~$i$  &Inclination (degrees)\dotfill & \ib  \\
\noalign{\smallskip}
                ~~~$e$ \tablefootmark{a}  &Eccentricity \dotfill & 0      \\
\noalign{\smallskip}                                      
                ~~~$R_{P}/R_\star$ &Ratio of  planet radius to stellar radii\dotfill & \rrb \\
\noalign{\smallskip}                
                ~~~$a/R_\star$  & Ratio of semi-major axis to stellar radii\dotfill & \arb   \\
                \noalign{\smallskip}                                              
                  ~~~$a$ &Semi-major axis (AU)\dotfill & \ab   \\
\noalign{\smallskip}                
                ~~~$u_1$ &Linear limb-darkening coeff\dotfill & \uone    \\
\noalign{\smallskip}                              
                ~~~$u_2$  &Quadratic limb-darkening coeff\dotfill & \utwo        \\
                
\noalign{\smallskip}
\multicolumn{3}{l}{\emph{RV Parameters}}\\
\noalign{\smallskip}
                ~~~$K\tablefootmark{b}$ &RV semi-amplitude variation (\ms)\dotfill &\kb     \\
\noalign{\smallskip}
                ~~~$K\tablefootmark{c}$ &RV semi-amplitude variation (\ms)\dotfill &\kbfco  \\
                                
\noalign{\smallskip}
\multicolumn{3}{l}{\emph{Planetary parameters}}\\
\noalign{\smallskip}                
                   ~~~$R_{P}$ & Planet radius (R$_\oplus$)\dotfill & \rpb \\
 \noalign{\smallskip}                                             
                  ~~~$M_{P}\tablefootmark{b}$ & Planet mass (M$_\oplus$)\dotfill   & \mpbGP  \\
\noalign{\smallskip}                                             
                  ~~~$M_{P}\tablefootmark{c}$ & Planet mass (M$_\oplus$)\dotfill   & \mpb  \\
\noalign{\smallskip}                                              
                  ~~~$\rho_\mathrm{p}\tablefootmark{b}$ & Planet density (g\,cm$^{-3}$) \dotfill &  \denpb   \\                  
\noalign{\smallskip}                                              
                  ~~~$\rho_\mathrm{p}\tablefootmark{c}$ & Planet density (g\,cm$^{-3}$) \dotfill &  \denpbfco   \\
\noalign{\smallskip}                
           ~~~$F$ &Insolation  ($F_\mathrm{\oplus}$)\dotfill & \Fequib \\

\noalign{\smallskip}                
           ~~~$T_{eq}\,\tablefootmark{d}$ &Equilibrium temperature (K)\dotfill & \Tequib \\
           
 \noalign{\smallskip}                
           ~~~$\Lambda\,\tablefootmark{e}$ &Restricted Jeans escape parameter \dotfill & $\approx$\,\Lambdap \\
       
\noalign{\smallskip}                
\hline
\end{tabular}
\label{Table: Orbital and planetary parameters}
\tablefoot{
\tablefoottext{a}{Fixed value.}
\tablefoottext{b}{Derived using a GP regression method.}
\tablefoottext{c}{Derived using the FCO technique.}
\tablefoottext{d}{Assuming isotropic re-radiation,  and a Bond albedo of zero.}
\tablefoottext{e}{Defined in \citet{2017A&A...598A..90F}.}
}
\end{table*}

\subsection{Floating chunk offset technique} \label{Subsect: Floating chunk offset method}
It is difficult to remove the influence of activity from RV measurements in a reliable way, particularly
for sparse data. The GP method often gives good results, but in our case it is trained using the
\emph{K2}  light curve that was taken before the RV measurements. Possibly at that time the activity signal 
could have shown different characteristics. It is therefore important to use independent techniques,
when possible, to determine the $K$ amplitude of the orbit.

The FCO technique is another method for filtering out the effects of activity,
but in a model independent way \citep{2014A&A...568A..84H}. Basically, it fits a Keplerian orbit to RV data that have
been divided into small time chunks, keeping the period fixed, but allowing the zero point offsets
to  float. The only assumption of the method is that the orbital period of the planet
is less than the rotational period of the star or other planets. The RV variations in one time chunk
is predominantly due to the orbital motion of the planet and all other variations 
constant. This method also naturally accounts for different velocity offsets between different instruments
or night-to-night systematic errors. As long as the timescales for these are shorter than the orbital
period, their effects are absorbed in the calculation of the offset.

 \begin{figure}[!t]
 \centering
   \includegraphics[width=11.8cm, angle=-90, scale=0.75]{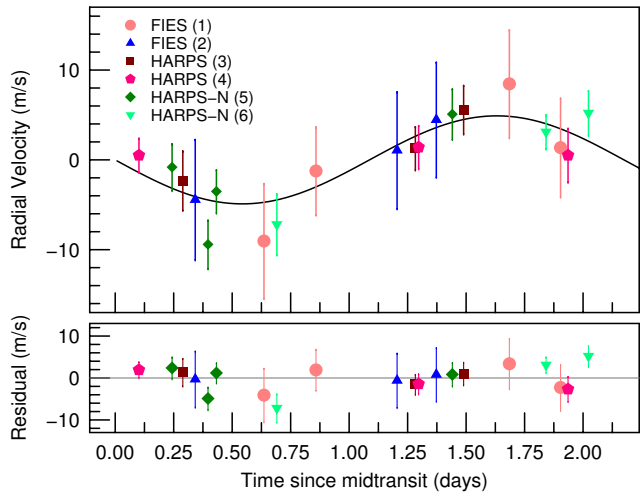}
   \caption{Radial velocity curve   phase folded to the orbital period of the planet (2.17~days)   
   using the FCO technique.  The RV data from the different chunks from each spectrometer are indicated in 
   a variety of colours and symbols. The double point used in both chunk three and four is plotted twice with a slight phase offset
   for clarity. 
   Seven of the time isolated  RVs  have been removed from the fit. 
   The resulting $K$ amplitude is \kbfco~m~s$^{-1}$.
   The lower panel shows residuals of the fit.}
      \label{Figure: FCO RV phase curve}
 \end{figure}

The FCO method is usually applied to ultra-short period planets ($P_\mathrm{orb} < 1$~day),
where the orbital motion in one night can be significant \citep[see][]{2014A&A...568A..84H}. However,
it can also be applied for planets on longer period orbits as long as these are shorter
than say, the rotational period of the star. One also should have relatively high cadence measurements. 
In the case of \target, the orbital period of the planet is 2.17~days and the best estimate
of the rotational period of the star is $\approx30$~days. Furthermore, we have high cadence 
measurement where observations were taken on several consecutive nights. The conditions
are right for applying the FCO method.

The data were divided into six data sets or chunks. It is important to exclude isolated measurements,
separated by more than several orbital periods, as these do not provide any shape information
for the RV curve. 
We divided the RV data into six time chunks that were separated 
by no more than two days 
with the exception of chunk one and four, which covered time spans 
of three and four days, respectively. 
In particular, the HARPS data were divided into two chunks of which the last had only two 
RV measurements. In order to include these data points, but to 
have more shape information,  the last RV value 
for chunk three was repeated in chunk four (and thus the time span for chunk four
increased to five days). The seventh column in Table~\ref{Table: RV measurements}  
 shows the division of the
RV chunks.

We first checked if the planet signal was present in our data
using the so-called FCO periodogram \citep{2014A&A...568A..84H}.
For this, the RV chunks are fit using a different trial period. The resulting $\chi^2$ as a function of period is a form
of a periodogram, and the $\chi^2$ should be minimised for the period that is present in the data.
This was carried out with trial periods spanning 0.5 -- 10~days. The reduced $\chi^2$ was minimised for a
period of 2.17~days as shown in   Fig.~\ref{Figure: FCO periodogram} (appendix~\ref{Sect: appendix figures}). 
This confirms that the RV variations due to the planet are clearly seen in our
data. 

An orbital fit was then made to the chunk data using the program Gaussfit \citep{1988CeMec..41...39J}.
The period and ephemeris were  fixed to the transit values, but the zero point offsets for each chunk and
the $K$ amplitude were allowed to vary. The resulting $K$ amplitude is 
\kbfco~m~s$^{-1}$, which
corresponds to a  
planet  mass of 
\mbox{\mpb~\mearth} (Table~\ref{Table: Orbital and planetary parameters}). The precision in mass
is 20~\%.  If we remove the double point in chunk four we get essentially the same amplitude 
($K = 5.1\pm1.0$~m~s$^{-1}$). 
Figure~\ref{Figure: FCO RV phase curve} 
shows the phased orbit fit after applying the calculated offsets. Different symbols indicate
the different chunks. This velocity amplitude is  in very good agreement with the GP analysis. 
The very small differences  merely reflect the variations due to a different
treatment of the activity signal.

When using the FCO method it is important to check that it can reliably recover an input $K$ amplitude.
The time sampling of the data or harmonics of the rotational period (e.g. \mbox{$P_\mathrm{rot}/2 \approx 15$~days})
may effect the recovered $K$ amplitude in a systematic way. This was explored through simulations. 
We first tried to account for any activity signal in a way independent from the GP model. To do this
we placed all the data on the same zero-point scale to account for the large relative offset between
the HARPS and FIES data and then removed the planet signal. A Fourier analysis showed no significant
peaks in the amplitude spectrum, but a weak one at 15~days with an amplitude of 3.5~m~s$^{-1}$. Assuming
this could be the first harmonic of the rotational period we fit a sine wave to the data using this
period and amplitude and took this as our activity signal. We note that a 15-day activity signal
should have a much larger effect on the results of the FCO method.
We then added the orbital signal of the planet to this activity signal using a range of $K$-amplitudes. 
The median error of our RV measurements is 2.8~\ms~so we added random noise with
$\sigma=3$~\ms. We also added a large velocity offset \mbox{($\approx$ $-$26 km\,s$^{-1}$)} between
the simulated FIES and HARPS/N measurements. Finally, for good measure we added an additional random
velocity component ranging between $-10$ to $+8$~\ms~to the individual chunks to account
for any additional activity jitter. For each input $K$-amplitude a total of 50 sample data sets were generated
using different random noise generated with different seed values. 
The mean and standard deviations were calculated for each. The $K$-amplitude was 
reliably recovered in the full amplitude range 1 -- 6~\ms. Figure~\ref{Figure: K amplitude FCO} 
(appendix~\ref{Sect: appendix figures}) shows
the  output $K$ amplitude as  a function of
input $K$ amplitude.   The red square is the value for \target.


\section{Discussion} \label{Section: Discussion}
\begin{figure*}[!t]
 \centering
            {\includegraphics[scale=0.45]{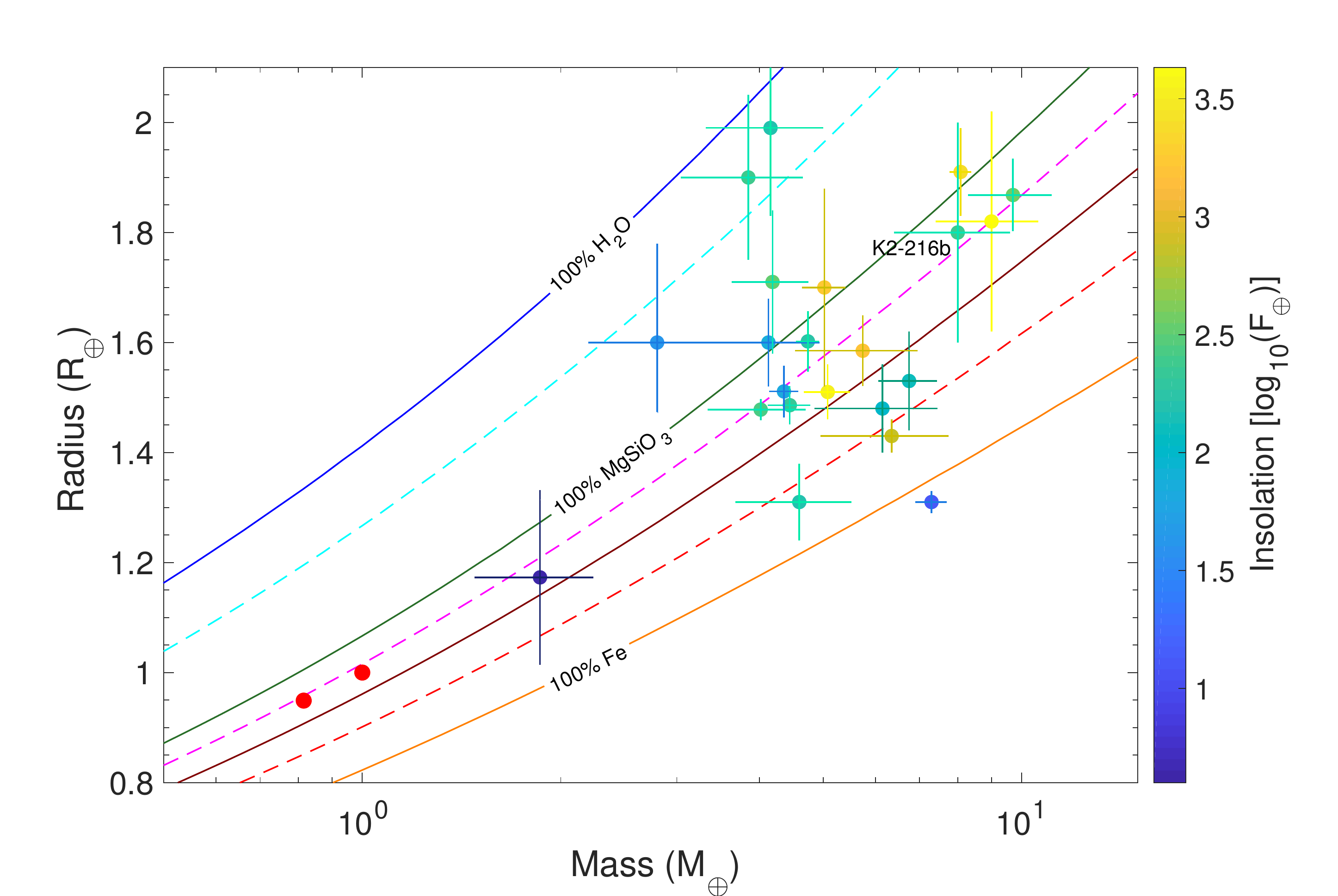}}  
   \caption{Mass-radius diagram of 
   all small exoplanets ($R_\mathrm{p} \leq 2$~\rearth, and $M_\mathrm{p} \leq 30$~\mearth) with a 
   measured mass and radius to a precision better than 20~\%   
 as listed in the NASA Exoplanet Archive. 
The colours of the planets indicate the insolation in units of 
$\log10\,(\mathrm{F}_\oplus)$.   
   Earth and Venus are plotted in red filled circles for comparison. 
 The solid lines are theoretical mass-radius curves      \citep{2016ApJ...819..127Z}, from top to bottom: 
100~\%~H$_2$O (blue solid line),   
a mixture of 50~\%~H$_2$O and 50~\%~MgSiO$_3$ (cyan dashed line),   
100~\%~MgSiO$_3$ (green solid line), 
a mixture of 75~\%~MgSiO$_3$ and 25~\%~Fe  (magenta dashed line), 
 a mixture of  50~\%~MgSiO$_3$ and 50~\%~Fe  (brown solid line), 
 a mixture of  25~\%~MgSiO$_3$ and 75~\%~Fe  (red dashed line), 
 and  100~\%~Fe (orange solid line). 
   }
      \label{Figure: mass-radius diagram}
 \end{figure*}

Combining our  mass   and radius estimates of \target b, we find  
mean   densities of \mbox{\denpb~\gc}~and \mbox{\denpbfco~\gc}~from the GP and FCO methods, respectively, 
 in excellent agreement with each other. 
In Fig.~\ref{Figure: mass-radius diagram} we show the position of planet~b  on a mass-radius diagram  (FCO mass) 
compared to all small exoplanets \mbox{($R_\mathrm{p}  \leq 2$~\rearth) }with   masses \mbox{$\leq 30$~\mearth}~known to   
better than 20~\%, as listed in the NASA Exoplanet Archive. 
The insolation flux of the planets is  colour coded. 
The figure also displays the \citet{2016ApJ...819..127Z} theoretical 
 models of  planet composition 
 in different colours from 100~\% water  to 100~\% iron.   
 The density of \target b is consistent with a rocky   composition of primarily iron and magnesium silicate.

 The radius of \target b puts it in the  middle, or just below the lower edge, of the bimodal radius distribution of small planets 
 \citep{2017AJ....154..109F}, 
    using the  
  location and shape of the radius gap as  estimated 
 by   \citet{2017arXiv171005398V} with  
\begin{equation}
\log (R) = m \times \log(P) + a\ , 
\end{equation} 
where $m = -0.09^{+0.02}_{-0.04}$ and $a = 0.37^{+0.04}_{-0.02}$.  
For a period of 2.17~days,  the location of the centre of the valley is  around 2.2~\rearth. 
This suggests that \target b is a remnant core, stripped of its atmosphere.

%


To estimate the likelihood of \target b having an extended atmosphere, we begin by considering 
that during the early phases of planet evolution, when a planet comes out of the proto-planetary nebula, 
it goes through a phase of extreme thermal Jeans escape, the so-called boil-off \citep{2017ApJ...847...29O}. 
After this phase, the planet arrives at a more stable configuration in which the escape is driven by the stellar extreme-ultraviolet (XUV) flux 
\citep{2017A&A...598A..90F}. Whether a planet lies in the boil-off regime or not, can be determined on the 
basis of the restricted Jeans escape parameter, $\Lambda$, which is defined as \citep{2017A&A...598A..90F}
\begin{equation}
\Lambda = \frac{ G M_\mathrm{p} m_\mathrm{H}}{k_\mathrm{B} T_\mathrm{eq} R_\mathrm{p}}\ ,
\end{equation}
where $G$ is the gravitational constant, $m_\mathrm{H}$ is the hydrogen mass, 
$k_\mathrm{B}$ is the Boltzmann constant,  and $T_\mathrm{eq}$  
is the equilibrium temperature of the planet. When \mbox{$\Lambda \leq 20 - 40$}, 
depending on the system parameters, a planet with a hydrogen-dominated atmosphere 
will lie in the boil-off regime \citep{2017A&A...598A..90F}. Considering the two derived 
planetary masses for \target b, $\Lambda$ ranges between   29 and 31. Assuming 
that the planet originally accreted a hydrogen-dominated atmosphere, following the boil-off 
phase, the planet will have a $\Lambda$ value of about 20 (to be conservative). This value corresponds  
to a planetary radius of \mbox{$\approx2.6$~\rearth}. The right panel in Fig.~4  of \citet{2011ApJ...738...59R} 
indicates that, following the boil-off phase \target b had a hydrogen-dominated atmosphere with 
a mass of \mbox{$\approx0.1$~\%} planetary mass.

 To examine whether this atmosphere would have escaped within the age of the system under 
the action of the high-energy (X-ray and EUV; XUV) stellar irradiation, we computed upper atmosphere 
models with the derived planet parameters employing the hydrodynamic code described by 
\citet{2018A&A...612A..25K}. We estimated the stellar XUV flux starting from the $\log\,(R^\prime_{HK})$ 
value derived from the spectra. We converted the measured $\log\,(R^\prime_{HK})$ value into 
\mbox{\ion{Ca}{ii}~H~\&~K} line core emission flux at 1~AU employing the equations listed in 
\citet{2017A&A...601A.104F}, obtaining 18~erg~cm$^{-2}$~s$^{-1}$. From this value, and using 
the relations given by \citet{2013ApJ...766...69L, 2014ApJ...780...61L}, we obtained a stellar 
Ly\,$\alpha$ flux at 1~AU of \mbox{20~erg~cm$^{-2}$~s$^{-1}$} and an XUV flux at the planetary 
orbit of approximately \mbox{19\,000~erg~cm$^{-2}$~s$^{-1}$}. Inserting this value and the planet 
parameters in the upper atmosphere code leads to mass-loss rates of 
\mbox{$6-9\times10^{-12}$~\mearth~year$^{-1}$}. This implies that the planet must have 
lost between 0.07 and 0.13\% of its mass in one Gyr.  
Our estimated age of the system in Sect.~\ref{Subsection: stellar mass and radius} has very large uncertainties,
but  during a 5 Gyr main-sequence lifetime of the host star the planet would have lost 
between 0.35 and 0.65\% of its mass, which is significantly larger than the 
predicted initial hydrogen-dominated envelope mass of 0.1\%. In addition, the above mass-loss 
predictions should be considered to be lower limits, since the young star was 
significantly more active than taken into account above. Thus, even considering the large uncertainties
in age, we   conclude that \target b likely has 
completely lost its primordial, hydrogen-dominated atmosphere, and is one of the 
largest planets found to have lost its atmosphere \citep[see Fig.~7 in][]{2017arXiv171005398V}.

\section{Summary}\label{Section: Summary}
In this paper, we confirm  the discovery  of  the super-Earth \target b (\targetaa b)   in a 2.17-day orbit
transiting a moderately active \stype~star at a distance of \distancegaia~pc. 
We derive the mass of planet b using two different methods: first, a GP regression based on both the RV and photometric time series, 
and second, the FCO  technique, based   on RV measurements 
observed close in time   with the assumption that the orbital period of the planet is much less than the
rotational period of the star.
The results are in very good agreement with each other:   
\mbox{\mplanet~$\approx$~\mpbGP~\mearth} from the GP regression and   
    \mbox{\mplanet~$\approx$~\mpb~\mearth} from the FCO technique. 
The density is consistent with a 
 rocky   composition of primarily iron and magnesium silicate, 
 although the   uncertainties   
allow a   range of planetary compositions. 
 With a size of \rpb~\rearth, this planet falls within, or just below, the gap of the bimodal radius distribution where
few planets are found.   Our results indicate that the planet has completely lost its primordial
hydrogen-dominated atmosphere,  supporting the formation scenario of short-period super-Earths 
as remnants from mini-Neptune planets.

\begin{acknowledgements}
We thank the  NOT, TNG, ESO,  Subaru, and TCS staff members for their  support during the 
observations. 
Based on observations obtained with (a) the Nordic Optical Telescope (NOT), operated on 
the island of La Palma jointly by Denmark, Finland, Iceland, Norway, and Sweden, in the Spanish Observatorio 
del Roque de los Muchachos (ORM) of the Instituto de Astrod\'isica de Canarias (IAC) (programmes 53-016, 54-027, and 
54-211); 
b) with the Italian Telescopio Nazionale Galileo (TNG)  operated at the ORM (IAC) on 
the island of La Palma by the INAF Fundaci\'on Galileo Galilei (programmes CAT16B\_61, CAT17A\_91, A36TAC\_12,
and OPT17B\_59);
(c) the 3.6 m ESO telescope at La Silla Observatory (programmes 
098.C-0860, 099.C-0491, and 0100.C-0808);   
(d) the Telescopio Carlos S\'anchez (TCS) installed at  IAC's   Observatorio del Teide, Tenerife; 
(e) the Subaru Telescope, operated by the National Astronomical Observatory of Japan; 
(f) NESSI,  funded by the NASA Exoplanet Exploration Program and the NASA Ames Research Center. 
NESSI was built at the Ames Research Center by Steve B. Howell, Nic Scott, Elliott P. Horch, and Emmett Quigley; 
(g) the \emph{K2/Kepler} mission. Funding for the \emph{K2/Kepler}  mission is provided
by the NASA Science Mission Directorate. 
The \emph{K2}  data  presented in this paper were downloaded from the Mikulski Archive for Space Telescopes 
(MAST). 
STScI is operated by the Association of Universities for Research in Astronomy, Inc., under NASA contract NAS5-26555. 
Support for MAST for non-HST data is provided by the NASA Office of Space Science via grant NNX13AC07G and by other grants and contracts.
This work 
has made use of SME package, which benefits from the continuing development work by J. Valenti and N. Piskunov 
and we gratefully acknowledge their continued support. This work has made use of the VALD database, operated 
at Uppsala University, the Institute of Astronomy RAS in Moscow, and the University of Vienna 
\citep{Kupka2000, Ryabchikova2015}.  
C.M.P. and M.F. gratefully acknowledge the support of the 
Swedish National Space Board.  DG gratefully acknowledges the financial support of the  Programma Giovani Ricercatori -- Rita Levi Montalcini -- Rientro dei Cervelli (2012)  awarded by the Italian Ministry of Education, Universities and Research (MIUR). 
DK and LF acknowledge the Austrian Forschungs-f\"orderungsgesellschaft FFG project ``TAPAS4CHEOPS'' P853993. 
SzCs, APH, MP, and HR acknowledge the support of the DFG priority programme
SPP 1992 "Exploring the Diversity of Extrasolar Planets (HA 3279/12-1,
PA525/18-1, PA525/19-1, PA525/20-1 and RA 714/14-1). Funding for the Stellar Astrophysics Centre is provided by The Danish 
National Research Foundation (Grant agreement no.: DNRF106). 
This project has received funding from the European Union's Horizon 2020 research and innovation programme 
under grant agreement No 730890. This material reflects only the authors views and the Commission is not liable 
for any use that may be made of the information contained therein.
We  thank the anonymous referee  whose constructive comments led to an improvement of the paper. 
 \end{acknowledgements}

\bibliographystyle{aa}
\bibliography{references}

\appendix

\section{Figures}\label{Sect: appendix figures}

  \begin{figure*}[!ht]
 \centering
  \resizebox{\hsize}{!}
            {\includegraphics{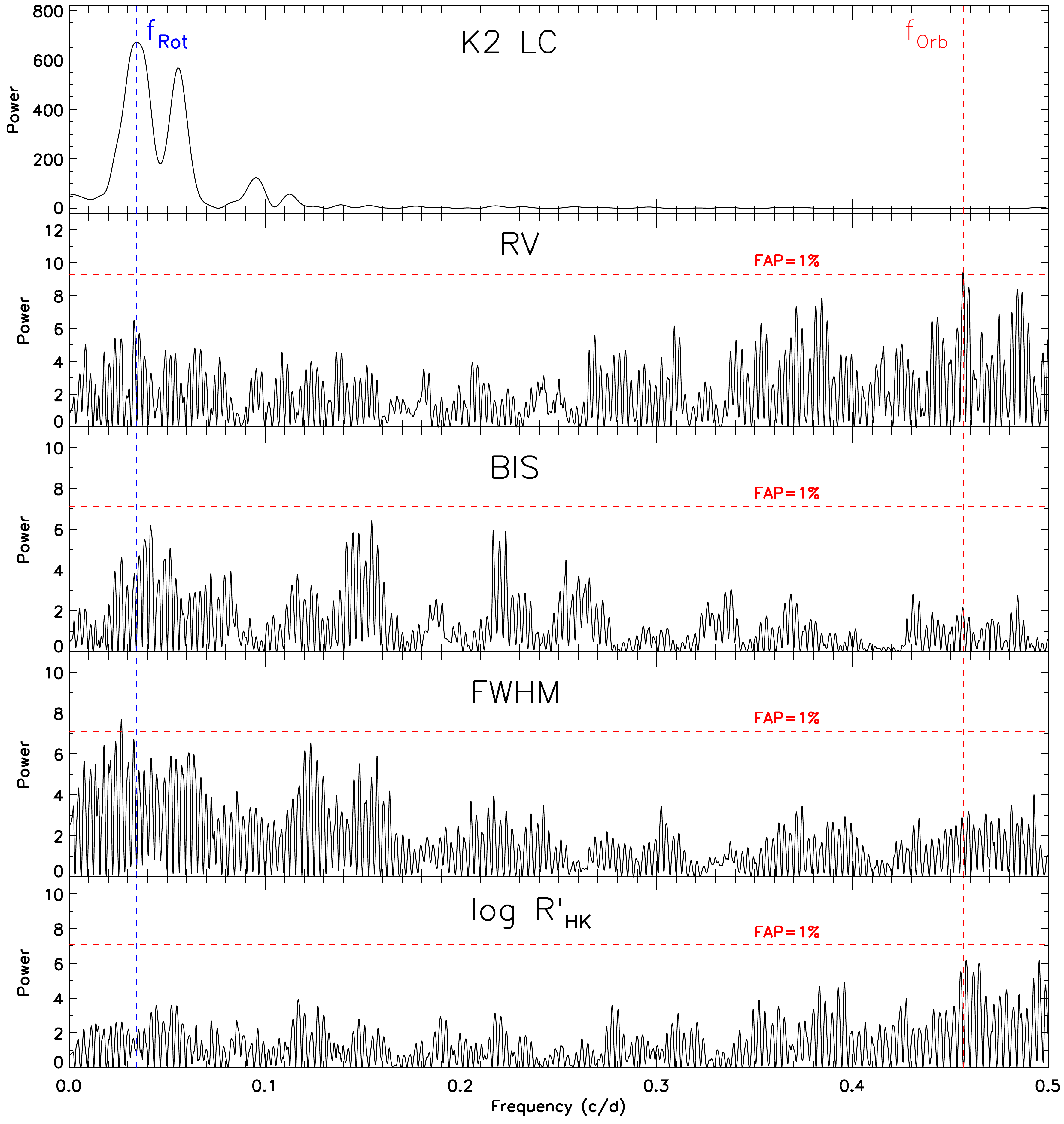}}  
   \caption{From top to bottom: Generalised Lomb-Scargle periodograms of the \emph{K2} light curve,
   the RV measurements, BIS, FWFM of the correlation function, and the activity index \lgr~where the last
   four are extracted from the
   FIES, HARPS, and HARPS-N data. The stellar period is indicated with the vertical dashed blue line, and
   the planet orbital period with the vertical dashed red line. The false-alarm probability (FAP) is indicated
   at the \mbox{1\% level}.}
      \label{Figure: Davide periodograms}
 \end{figure*}
 \begin{figure}[!t]
 \centering
  \resizebox{\hsize}{!}{
   \includegraphics[angle=-90]{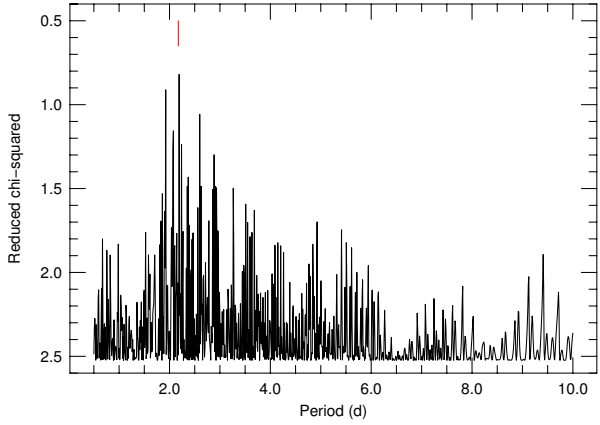}}  
   \caption{Floating chunk offset-periodogram over the range $0.5 - 10$~days ($\chi^2$
vs. period).  The y-axis is flipped so that a minimum
appears as  a peak, much like a standard periodogram.  
The best fit is at the planet period of 2.17~days.}
      \label{Figure: FCO periodogram}
 \end{figure}

 \begin{figure}[!t]
 \centering
  \resizebox{\hsize}{!}{
   \includegraphics[angle=-90]{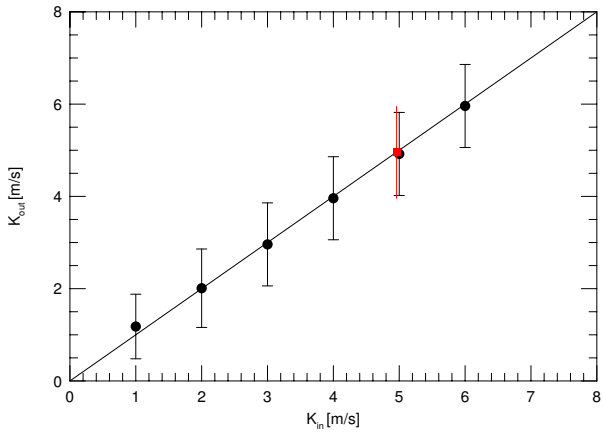}}   
   \caption{Output $K$ amplitude as  a function of
input $K$ amplitude using the FCO technique.}
      \label{Figure: K amplitude FCO}
 \end{figure}

\end{document}